\documentclass[journal=jacsat,manuscript=article]{achemso}

\usepackage[version=3]{mhchem} 
\usepackage{multirow}
\usepackage{makecell}
\usepackage{cellspace}
\usepackage{booktabs}
\usepackage{amsmath} 
\usepackage{amssymb} 

\author{Janghoon Ock}
\author{Srivathsan Badrinarayanan}
\affiliation[ChemE]
{Department of Chemical Engineering, Carnegie Mellon University, 5000 Forbes Street, Pittsburgh, PA 15213, USA}
\author{Rishikesh Magar}
\author{Akshay Antony}
\author{Amir Barati Farimani}
\email{barati@cmu.edu}
\affiliation[MechE]
{Department of Mechanical Engineering, Carnegie Mellon University, 5000 Forbes Street, Pittsburgh, PA 15213, USA}

\title[An \textsf{achemso} demo]
  {Multimodal Language and Graph Learning of Adsorption Configuration in Catalysis}

\abbreviations{IR,NMR,UV}

\begin{document}
\begin{tocentry}
\centering
\includegraphics{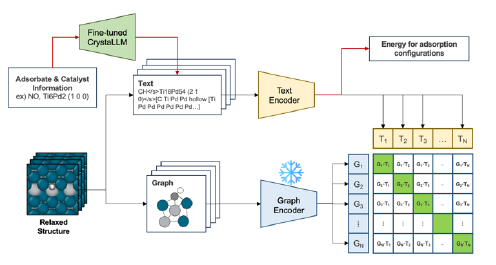} 
\label{fig:toc}

\end{tocentry}

\begin{abstract}
Adsorption energy is a reactivity descriptor that must be accurately predicted for effective machine learning (ML) application in catalyst screening. This process involves finding the lowest energy among different adsorption configurations on a catalytic surface, which often have very similar energies. While graph neural networks (GNNs) have shown great success in computing the energy of catalyst systems, they rely heavily on atomic spatial coordinates. In contrast, transformer-based language models can directly use human-readable text inputs, potentially bypassing the need for detailed atomic positions or topology. However, these language models often struggle with accurately predicting the energy of adsorption configurations. Our study improves the predictive language model by aligning its latent space with well-established graph neural networks (GNNs) through a self-supervised process called graph-assisted pretraining. This method reduces the MAE of energy prediction for adsorption configurations by 7.4-9.8\%, redirecting the model's attention toward adsorption configuration. Building on this, we propose using generative large language models to create text inputs for the predictive model without relying on exact atomic positions. This demonstrates a potential use case of language models in energy prediction without detailed geometric information.
\end{abstract}


\section{Introduction}
Machine learning (ML) approaches, particularly Graph Neural Networks (GNNs), have emerged as efficient surrogates to computationally expensive Density Functional Theory (DFT) simulations \cite{mlp, OC20_intro, OC20,Reiser2022}. These advancements can accelerate energy and force predictions for high-throughput material screening \cite{MLforCat,  Brook2022, Tran2022, adsorbml, ml_membrane}. The successful application of ML-based DFT surrogate modeling in catalysis can enable the identification of optimal catalyst materials for specific reactions, which is crucial for advancing energy storage technologies and sustainable chemical processes. The importance of such techniques has drawn attention beyond the chemical engineering and chemistry communities, extending into the AI for Science field \cite{opencatalyst2023}.

Despite the significant success of GNNs in machine learning applications in the catalysis domain, obtaining their input data can be challenging since they require atomic positions or topology. Constructing graph representations of structures relies on identifying nearest neighbors within specific proximity thresholds for each atom\cite{cgcnn, schnet, gemnet-oc, reviewer3}. However, achieving such precise coordinates can be difficult, limiting the applicability of GNNs primarily to theoretical studies. For instance, even with experimentally validated adsorption energy data from the literature, using this information in modeling remains difficult because replicating the exact atomic positions of the adsorbate-catalyst systems from experiments is problematic\cite{challenge, challenge2}.

Recent advancements in language model applications offer a promising alternative to relying on exact atomic coordinates as input data \cite{moformer, gpt-molberta, transpolymer, catberta}. Language models can process human-readable text descriptions of atomic systems instead of building an input with atomic coordinates. For example, the MOFormer model encodes metal-organic frameworks (MOFs) as text string representations, called MOFid, which include chemical information on building blocks and topology codes, unlike graph representations \cite{moformer}. The TransPolymer model encodes polymers using the SMILES strings of their repeating units along with attributes such as the degree of polymerization, polydispersity, and chain conformation \cite{transpolymer}. Additionally, the ability to process textual input allows us to incorporate experimentally-obtainable attributes into the input data. We aim to extend these successes from the materials science domain to the catalysis domain. For instance, the CatBERTa model takes textual input for adsorbate-catalyst systems to predict the energy of the system\cite{catberta}.

Identification of adsorption energy is an important task in catalysis since it is a key reactivity descriptor in catalyst screening \cite{BEP, BEP2, OC20}. A single adsorbate-catalyst pair can have numerous adsorption configurations, varying by adsorption site and molecule orientation on the catalytic surface\cite{adsorbml, Ock2023}. The minimum energy among these configurations is considered the adsorption energy. Due to the subtle differences between these configurations, their energies can be very similar. Therefore, to accurately identify the adsorption energy, the model must be capable of distinguishing these subtle energy differences, which can range from 0.1 to 0.3 eV around the minimum energy\cite{adsorbml}. Although the language models offers the potential to bypass the need for exact atomic positions, which are critical for building graph representations in many models used for adsorbate-catalyst systems, its accuracy remains a concern. Improving the model’s accuracy is essential to effectively apply this text-based approach to adsorption configuration energy prediction tasks.

To address this challenge, our study introduces graph-assisted pretraining, a multimodal learning method that leverages graph modality to improve the prediction accuracy of the language model for adsorption configurations. Multimodal learning has already been successfully applied to materials science and chemistry to boost model performance\cite{moformer, Huang2024, multipeptide}. We aim to extend this success to the catalysis domain, particularly to enhance the predictive capability of language-based models in adsorption energy prediction. Graph-assisted pretraining transfers the structural knowledge captured in graph embeddings to text embeddings in a self-supervised manner. This transfer of knowledge from a learned embedding space to the language model will help in our application of adsorption configuration energy predictions.

Additionally, we aim to show a potential use case of the Large Language Model (LLM) in making predictions without relying on precise atomic positions. For this, we use the generative capabilities of LLMs to generate desired textual input data for our predictive language model for energy prediction. Language models' generative capabilities have recently shown success in structure generation for inorganic crystals\cite{crystallm, gruver2024meta}. In this study, we specifically fine-tune CrystaLLM to generate Crystallographic Information Files (CIFs) for relevant adsorbate-catalyst systems instead of inorganic crystals. Here, the fine-tuned CrystaLLM takes textual information about the chemical composition of the system, along with its surface orientation. Subsequently, we use the generated CIFs to derive the input string for our predictive language model. This method allows us to make energy predictions without knowing the full structure of the adsorbate-catalyst configurations.


\section{Results and discussion}
\subsection{Framework}

\begin{figure*}[htbp] 
\centering
\includegraphics[width=0.9\textwidth]{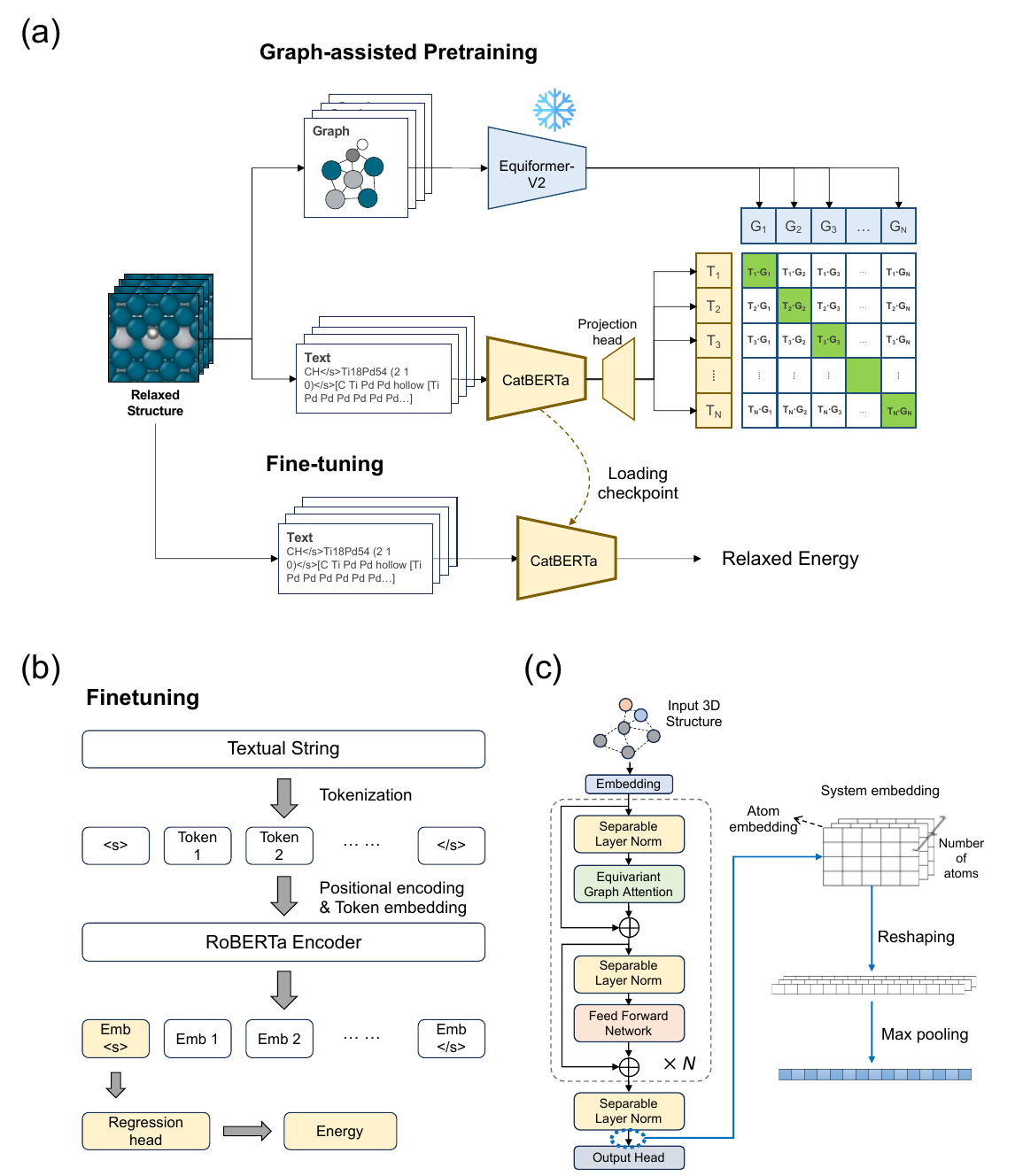} 
\caption{Overview of the model training framework. (a) The training process consists of two steps: graph-assisted pretraining and energy prediction fine-tuning. (b) The CatBERTa model is used as the text encoder. (c) The EquiformerV2 model serves as the graph encoder, and the graph embedding from the final layer is converted to a 1D format by reshaping and max pooling the collection of atom embeddings. The architecture image is reproduced from the original EquiformerV2 paper\cite{equiformerv2}.}
\label{fig:framework}
\end{figure*}

The language model-based approach for catalyst energy prediction leverages textual data for both training and inference. We have developed a multimodal pretraining framework, termed graph-assisted pretraining, to bridge the established graph-based approach with the newly introduced text-based approach within a shared latent space, as depicted in Figure \ref{fig:framework}. This method is introduced to enhance the accuracy of adsorption configuration energy predictions. This framework utilizes the CatBERTa model, which uses the RoBERTa encoder for text processing and a linear regression header to predict catalyst system energies (see Figure \ref{fig:framework} (b)) \cite{catberta}. Additionally, the EquiformerV2 model is employed as a graph encoder due to its capability to encode precise atomic structure (see Figure \ref{fig:framework} (c)) \cite{equiformerv2, opencatalyst2023}. In this framework, both text and graph embeddings are aligned in a self-supervised manner during pretraining. Subsequently, the model undergoes a fine-tuning stage, where it is trained in a supervised manner using energy labels derived from DFT calculations. Importantly, the fine-tuning step relies exclusively on text input data, without the need for graph representations. More details can be found in the Methods section.

\begin{figure*}[htbp] 
\centering
\includegraphics[width=0.9\textwidth]{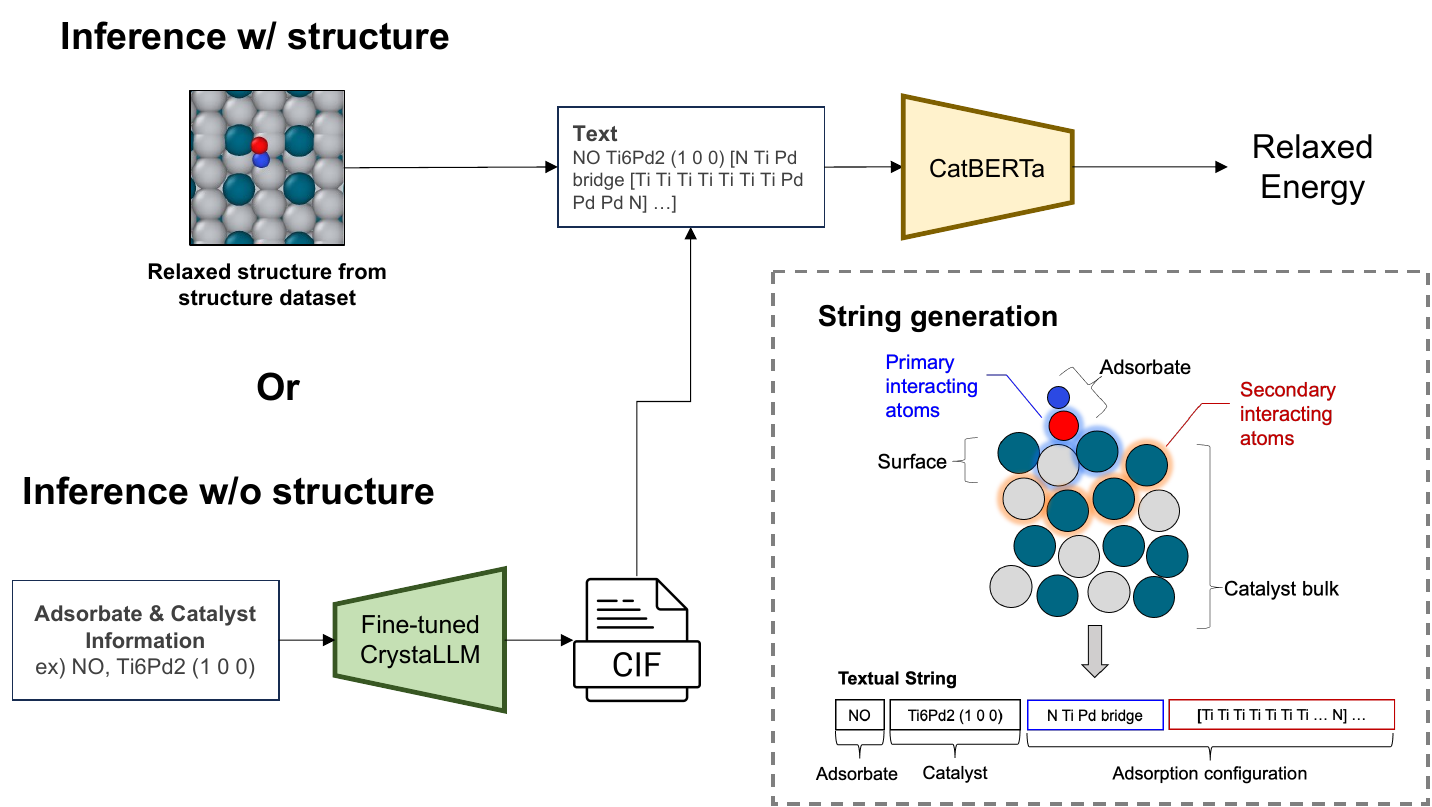} 
\caption{Model inference framework. Both structure data from the Open Catalyst datasets and CIFs generated by fine-tuned CrystaLLM can be converted into textual strings compatible with CatBERTa input, following the string conversion logic shown in the bottom right box. Generated CIFs provide structure information, including atomic positions, types, and unit cell details.}
\label{fig:inference}
\end{figure*}

We conduct two types of downstream inference: one to assess the effect of graph-assisted pretraining and the other to demonstrate the model's capability to predict energy without precise knowledge of the adsorbate-catalyst system structures. Both are depicted in Figure \ref{fig:inference}. First, to evaluate the impact of graph-assisted pretraining on prediction accuracy, we made predictions on the test set strings derived from the ML-relaxed structure. The CatBERTa model, which takes textual strings as input, is trained using textual data derived from ML-relaxed structures to predict the energy of a relaxed configuration. Second, to illustrate the model’s potential in predicting energies without relying on exact structures, we generate indicative structures in Crystallographic Information File (CIF) format using a large language model (LLM). This is done by providing the chemical composition and surface orientation of the adsorbate and catalyst as input. The generated CIFs are converted into textual strings compatible with CatBERTa input.

The textual strings are generated by converting structural information into a specific format containing three sections, as illustrated in the bottom box of Figure \ref{fig:inference}. The first section represents the adsorbate's chemical symbol, and the second part includes the catalyst's chemical symbols and Miller index, indicating the chemical composition and surface orientation respectively. The final section describes the adsorption configuration, capturing the primary and secondary interacting atoms in the adsorbate and the top layers of the catalyst surface, identified using the Pymatgen library \cite{pymatgen}. Further details on this structure-to-text conversion process are available in the Methods section.

\subsection{Data pipeline}

The textual string input for CatBERTa training is derived from the relaxed structures in the Open Catalyst 2020 (OC20) and Open Catalyst 2020 Dense (OC20-Dense) datasets. For both CatBERTa-involved training and CrystaLLM fine-tuning, training and validation are conducted using texts sourced from DFT-relaxed structures. Specifically, for the first case, we convert the relaxed structures to string representations and use them for the training and validation process. For the latter case, we create CIFs for the relaxed structures, then use them to fine-tune CrystaLLM.

In the first case of graph-assisted pretraining evaluation, predictions are made on strings generated from ML-relaxed structures. These ML-relaxed structures, along with their DFT-calculated energy labels, are provided by the Open Catalyst Project Challenge 2023\cite{opencatalyst2023}. GemNet-OC, SCN, and eSCN are used for the ML relaxation process. The ML-relaxations are conducted on out-of-domain splits in the OC20-Dense dataset, yielding 11,508, 11,630, and 11,755 relaxed structures from each model, respectively. To obtain valid DFT energies, DFT single-point calculations were performed on ML-relaxed structures. These were prepared by the dataset provider and included in the publicly available dataset. Our model's accuracy is then evaluated using approximately 920 of these ML-relaxed structures with valid DFT energies. We quantify the uncertainty of our model’s predictions by calculating the standard deviation across predictions for structures relaxed using GemNet-OC, SCN, and eSCN. Their individual results are listed in Supplementary Table S5. For embedding and attention score analysis, we use the entire set of ML-relaxed structures, ranging from 11,508 to 11,755, regardless of whether these structures have verified DFT energies. Details about the data split are shown in Supplementary Section S2.

For inference on the LLM-derived strings, predictions are made based on strings derived from adsorbate and catalyst,  as well as surface orientation. The aim is to demonstrate the potential of generating plausible textual string representations using the LLM framework. A subset of adsorbate and catalyst pair information is chosen from the original OC20-Dense training set, which contains 235 unique adsorbate-catalyst pairs. The downselection process is detailed in the Methods section. From these pairs, we extract only the adsorbate, catalyst, and Miller index information from these pairs and use them as initial prompts for the fine-tuned CrystaLLM framework.

\subsection{Graph-assisted pretraining}

Graph-assisted pretraining, a core component of our framework, is designed to transfer knowledge from graph embeddings to text embeddings. This approach bridges the gap between GNNs, which show great performance in energy and force predictions, and language models, which process human-interpretable data but do not take the entire structure as input. We select EquiformerV2 as the graph encoder due to its excellent performance with the OC20 dataset \cite{equiformer,equiformerv2}. The CatBERTa model serves as the text encoder, producing text embeddings that are then projected to match the dimensions of the graph embeddings. To align these embeddings, we apply a contrastive loss to increase the similarity between embeddings from the same adsorbate-catalyst configurations, with the graph encoder remaining frozen.

\begin{table}[htbp]
\centering
\caption{Performance comparison of CatBERTa with and without graph-assisted pretraining (GAP). ``Combined" refers to a combination of OC20 and OC20-Dense datasets.}
\label{tab:result}
\resizebox{\textwidth}{!}{%
\begin{tabular}{lcccccc}
\toprule
 & GAP Data (size) & Fine-tuning Data (size) & \multicolumn{2}{c}{Prediction Results} & \multicolumn{2}{c}{Improvement from GAP} \\
\cmidrule(lr){4-5} 
\cmidrule(lr){6-7}
 & & & MAE [eV] (\(\downarrow\)) & \( R^2 \) [-] (\(\uparrow\)) & MAE (\%) (\(\downarrow\)) & \( R^2 \) (\%) (\(\uparrow\))\\
\midrule
CatBERTa & - & OC20 (460k) & 0.713 ± 0.014 & 0.584 ± 0.014 & - & - \\
 & - & OC20-Dense (16k) & 0.542 ± 0.011 & 0.712 ± 0.008 & - & - \\
 & - & Combined (476k) & 0.378 ± 0.005  & 0.863 ± 0.005 & - & - \\
 \midrule
GAP-CatBERTa & OC20 (460k) & OC20 (460k) & 0.643 ± 0.020 & 0.691 ± 0.015 & -9.82 & +18.32 \\
 & OC20 (460k) & OC20-Dense (16k) & 0.502 ± 0.010 & 0.764 ± 0.008 & -7.38 & +7.30 \\
 & Combined (476k) & Combined (476k) & 0.346 ± 0.005 & 0.882 ± 0.002 & -8.47 & +2.20 \\
\bottomrule
\end{tabular}
}
\end{table}

Applying this method to the embedding space offers utility and flexibility. It operates solely on embeddings without downstream-task-specific labels, such as regression labels or classification categories. This means that we do not need to obtain labels for this pretraining stage. The downstream fine-tuning process can remain text-only, like the standard CatBERTa method. Once a properly pretrained checkpoint bridging graph and text modalities is established, it can be applied to multiple downstream tasks. Utilizing the embedding space also enhances generalizability, allowing the method to be applied to various encoders, provided their embedding sizes match. This graph-assisted pretraining method significantly improves prediction accuracy and adaptability across different datasets and tasks.

The graph-assisted pretraining method results in a substantial reduction in MAE, as shown in Table \ref{tab:result}, with decreases ranging from 7.4\% to 9.8\%. To evaluate the enhancement from graph-assisted pretraining, we compare the prediction results of CatBERTa with and without this pretraining method. In all cases, graph-assisted pretraining improves downstream prediction accuracy. 

Notably, pretraining with OC20 also benefits fine-tuning solely with OC20-Dense, despite there being no overlap between these datasets. This indicates that graph-assisted pretraining can serve as a transferable pretraining strategy, bridging the gap between high-performing GNNs and emerging Transformer-based language model approaches. It demonstrates the potential of self-supervised pretraining on one dataset to enhance performance on downstream tasks involving a different dataset. Prediction visualization and further analysis are provided in Supplementary Figure S2 and S3. While the significant accuracy improvement after incorporating OC20-Dense is attributed to the expansion of in-domain systems, what stands out is that a comparable level of improvement is still observed after applying graph-assisted pretraining.

\subsection{Enhancement in the latent space and attention score}

\begin{figure*}[h!] 
\centering
\includegraphics[width=0.95\textwidth]{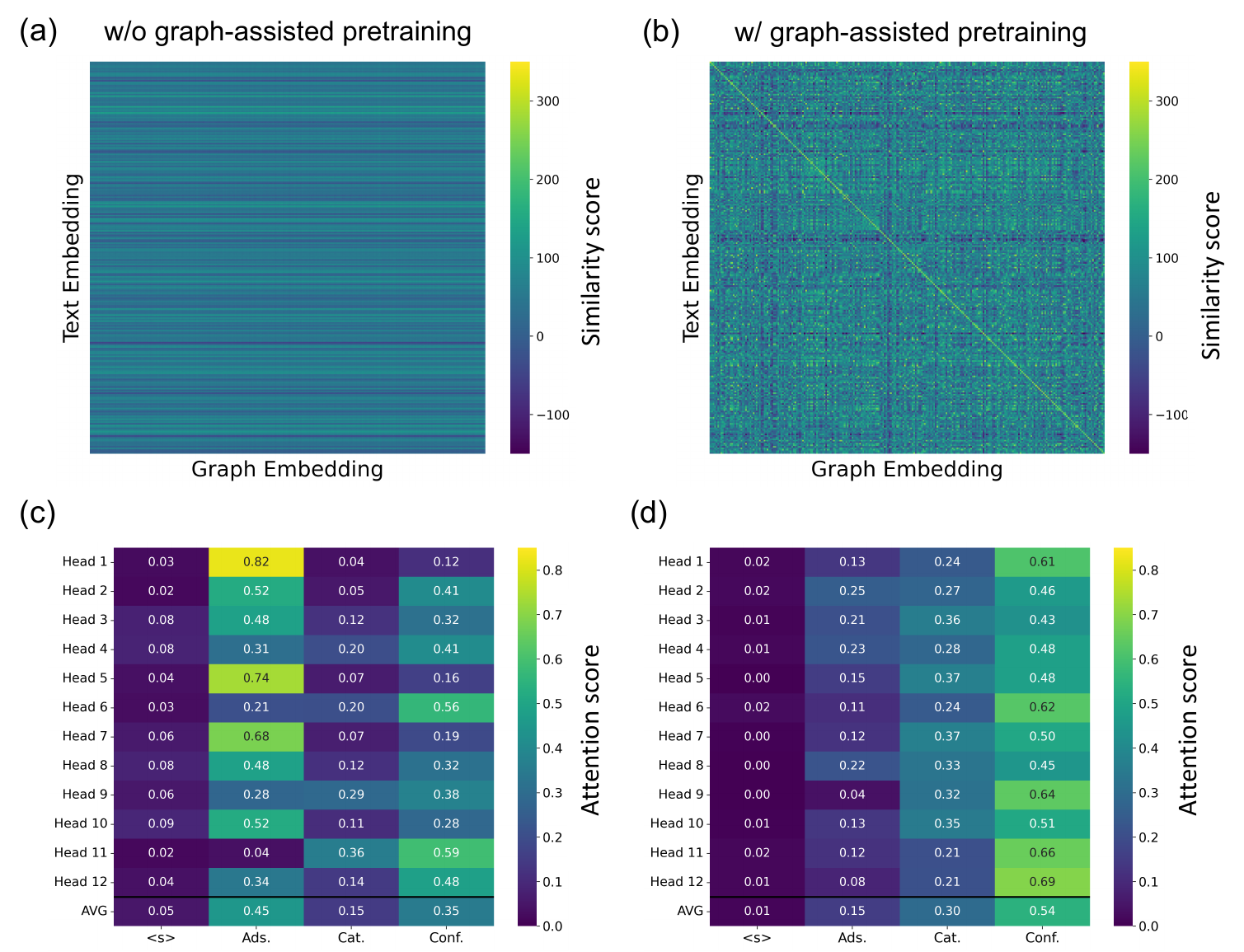} 
\caption{Analysis of similarity scores and sectional attention with and without graph-assisted pertaining. (a) and (b) displays similarity score analysis. (c) and (d) shows sectional attention score comparison. The left panels are without graph-assisted pretraining, while the right panels are with it. These results are derived from model predictions, which were trained on the OC20 dataset and evaluated using text strings from the GemNet-OC-relaxed structures.}
\label{fig:latent}
\end{figure*}

As graph-assisted pretraining is applied to the embeddings, it is essential to examine the latent space to observe its effects. Graph-assisted pretraining can align the graph and text embeddings from the same adsorbate-catalyst configurations. Figure \ref{fig:latent} (a) and (b) show the similarity matrix of graph and text embeddings. After applying graph-assisted pretraining, a clear diagonal line appears in the similarity matrix, indicating the alignment between embeddings in the latent space. 

Figure \ref{fig:latent} (a) and (b) show the similarity matrices of graph and text embeddings. After applying graph-assisted pretraining, a clear diagonal line appears, indicating the alignment between embeddings in the latent space. By comparing the left and right panels, we can clearly observe that the similarity score of actual pairs becomes higher than that of random pairs, which are not supposed to have correlations. The horizontal stripes in the similarity matrix before applying graph-assisted pretraining are due to the weight initialization in the final layer of EquiformerV2. Additionally, clustering in the latent space is shown in Supplementary Figure S1.

The analysis of attention scores in the final layer provides insights into how the model allocates attention. Our input string consists of three sections as discussed earlier, and the section-wise attention scores reveal the model's focus on each section. We extract and average the attention scores for the ``\textless{}s\textgreater{}" token, which is fed to the regression head, across three distinct sections: the adsorbate, the catalyst, and the adsorption configuration. Additionally, we compute the attention score of the ``\textless{}s\textgreater{}" token with respect to itself. The section-wise averaged attention scores are presented in Figure \ref{fig:latent} (c) and (d).

Graph-assisted pretraining redirects the model’s attention toward the adsorption configuration section. This attention redirection occurs across all 12 attention heads. Specifically, while the vanilla CatBERTa model primarily concentrates on the adsorbate section, the graph-assisted pretraining reallocates the model's focus towards both the catalyst and configuration sections. This shift in attention aligns with the physical principle that the interaction of the adsorbate with the catalytic surface as a whole is more critical than analyzing the adsorbate and the surface as separate entities\cite{Gao2020, Jacques2020}.

\subsection{Energy prediction for unknown structures}

One benefit of using language models and language representations is to bypass the need for explicit atomic structures. The CatBERTa model is capable of processing text descriptions with or without atomic position information. However, for optimal performance, the model still relies on incorporating neighbor atom information within the textual input, as demonstrated in previous predictions. To address this, we explore using an additional LLM to generate the necessary input data solely based on adsorbate and catalyst information. The CatBERTa input string contains three sections, as illustrated in Figure \ref{fig:inference} (a). The main idea is to derive the last configuration section from the first two sections, which pertain to the adsorbate and catalyst details.

\begin{figure*}[h!] 
\centering
\includegraphics[width=0.9\textwidth]{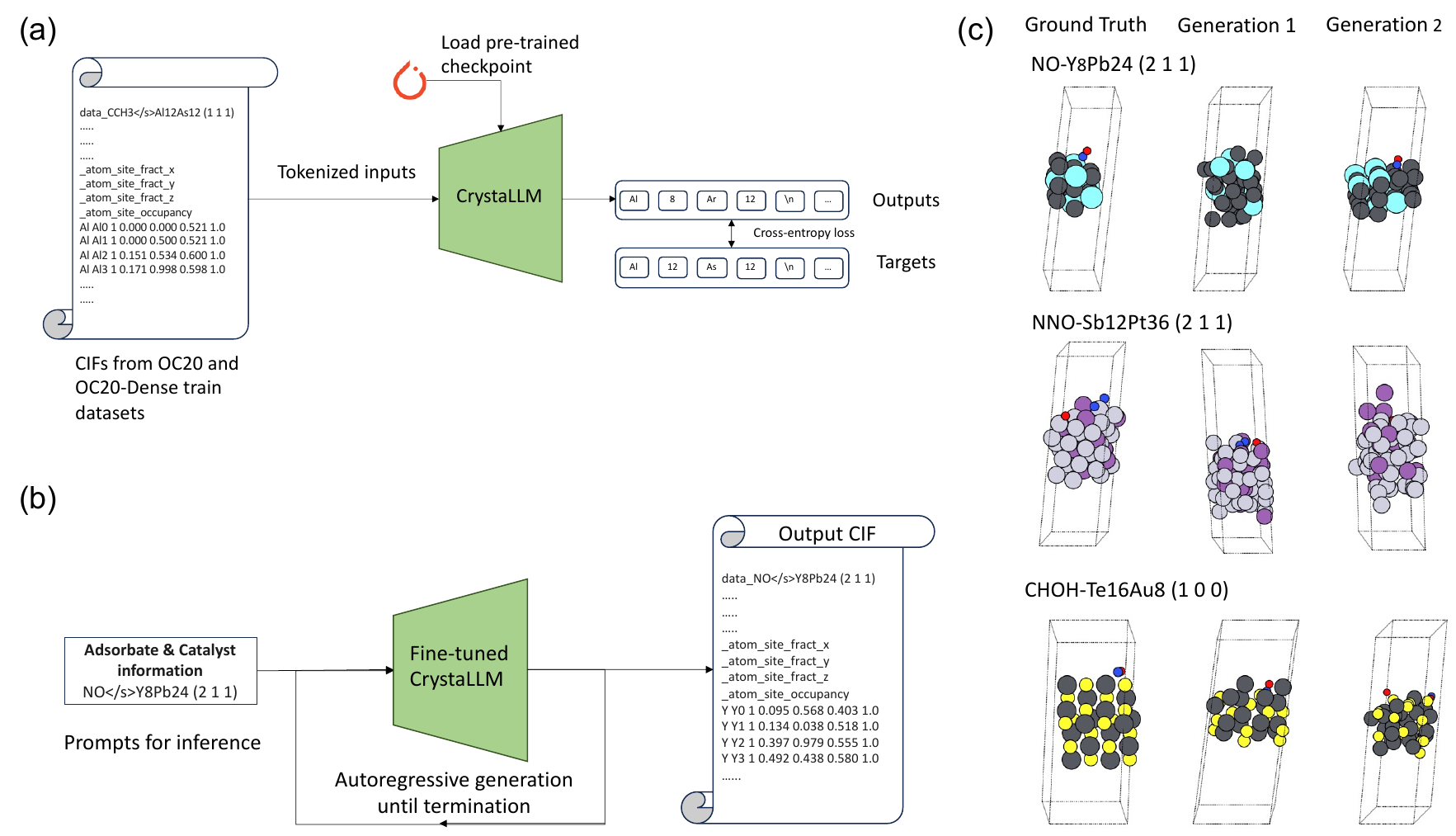} 
\caption{CrystaLLM framework. (a) illustrates the fine-tuning step using the CIFs from the relaxed structures in the OC20 and OC20-Dense training datasets. (b) depicts the inference process using the provided adsorbate and catalyst pair information. (c) shows visualization examples. These ground truth systems are sourced from the OC20 validation set, matching composition and surface orientation.}
\label{fig:crystallm}
\end{figure*}

For this purpose, we use CrystaLLM, which was originally trained to generate CIFs of inorganic crystals \cite{crystallm}. We fine-tune the pretrained CrystaLLM using CIFs from relaxed structures in the Open Catalyst training datasets, as illustrated in Figure \ref{fig:crystallm}. For details on the process and hyperparameters, please refer to the Methods section and Supplementary Table S2. CrystaLLM autoregressively predicts the next tokens in the CIF until it encounters two consecutive `\textbackslash n' tokens. The initial prompt to the CrystaLLM is set as the first two parts of the CatBERTa input string, which includes adsorbate and catalyst information along with the Miller index. These first two parts of the CatBERTa input string are derived from the metadata of the Open Catalyst dataset\cite{OC20, adsorbml}, encompassing the adsorbate chemical symbol, catalyst chemical symbol, and Miller index—information that relies on atomic geometry and is experimentally obtainable.

As the model autoregressively generates the next token based on the given tokens, the model completes the rest of the CIF from the given starting prompt. For example, the input for the fine-tuned CrystaLLM might be `data\_CCH3\textless{}/s\textgreater{}Al12As12 (1 1 1)', and the output from the model would be the corresponding CIF file, which contains indicative structure information. ``Indicative" means that, while the generated CIFs do not necessarily guarantee atomistic validity, they still provide some information about neighboring atoms that can assist in improving CatBERTa’s predictions (see Figure \ref{fig:crystallm} (c)). Since the generated CIFs lack precise atomic coordinates, they cannot be used as input for GNNs.

For unknown structure systems, we can use the proposed generative language model approach along with the predictive CatBERTa model to obtain energy predictions. From the OC20-Dense training dataset, which contains 235 unique adsorbate-catalyst pairs, we downsampled 66 pairs based on the type and number of elements in the adsorbate and catalyst bulk. These selected pairs of adsorbate and catalyst information are used as starting prompts, which are fed into the fine-tuned CrystaLLM to generate CIFs (see Figure \ref{fig:crystallm} (b)). We iterated through three generations and selected the CIFs where the composition of generated atoms matched the given adsorbate and catalyst chemical symbols within a certain threshold. The detailed process is provided in the Methods section. These CIFs are then converted into textual string inputs for CatBERTa prediction. Each adsorbate-catalyst pair can have numerous adsorption configurations in the OC20-Dense dataset. For the 66 downsampled adsorbate-catalyst pairs, the total number of adsorption configurations is 5,141.

\begin{figure*}[h!] 
\centering
\includegraphics[width=0.9\textwidth]{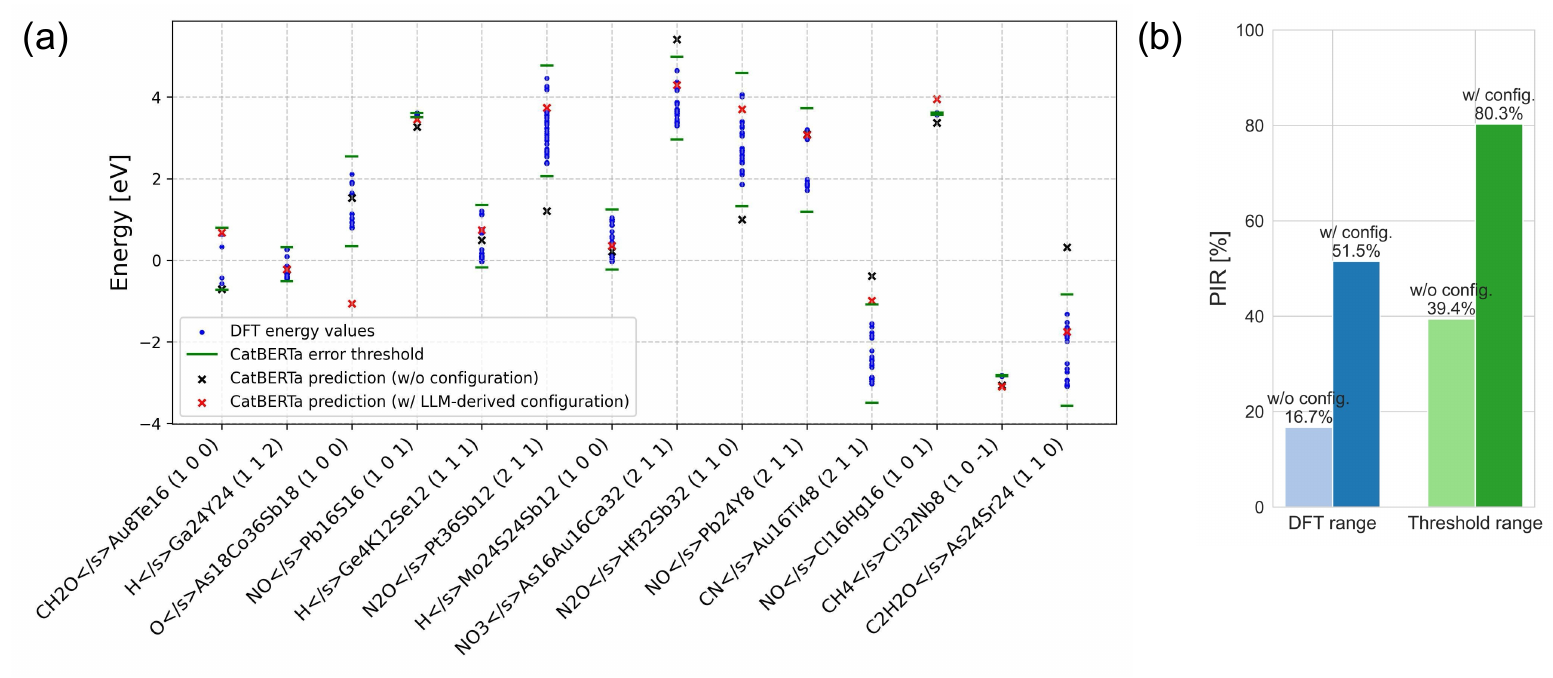} 
\caption{Enhancement from LLM-derived strings as input for the CatBERTa model. (a) Twelve example adsorbate-catalyst pairs are sampled from the 66 pairs. Blue dots represent the energy of different adsorption configurations for each adsorbate-catalyst pair. The number of adsorption configurations for the twelve example pairs ranges from 4 to 130, with a mean value of 62.5. (b) Prediction Inclusion Ratio (PIR) for each case quantifies the improvement in prediction accuracy across 66 pairs. The term `config.' refers to the LLM-derived configurations strings.}
\label{fig:energy_plot}
\end{figure*}

We benchmark the energy predictions using these adsorbate-catalyst pairs to determine if the LLM-derived configuration strings can help the energy prediction. For this, we compared two types of predictions; one is made only with the adsorbate and catalyst section, while the other is made on the strings including LLM-derived configuration strings. Subsequently, we compare those predicted values with the actual DFT energy values of possible adsorption configurations for those chosen adsorbate-catalyst pairs in the OC20-Dense dataset, as shown in Figure \ref{fig:energy_plot} (a). For this prediction, we used CatBERTa, pretrained and fine-tuned with the OC20 dataset, making these predictions entirely out-of-domain since the adsorbate-catalyst pairs are from the OC20-Dense dataset.

This comparison allows us to evaluate whether the LLM-derived strings can help predictions by bringing them within the plausible energy range across possible configurations. To quantify this, we define the Prediction Inclusion Ratio (PIR) as:

\begin{equation}
\text{PIR [\%]} = \frac{N_{\text{in-range}}}{N_{\text{total}}}
\end{equation}

where \( N_{\text{in-range}} \) represents the number of predictions falling within the desired range, and \( N_{\text{total}} \) is the total number of predictions.

In Figure \ref{fig:energy_plot} (a), the blue points represent the DFT-calculated energy values for various adsorption configurations of each adsorbate-catalyst pair. The red X marks represent CatBERTa predictions with LLM-derived configurations, while the black X marks represent predictions without them. The range of blue points indicates the potential variations in DFT energy of adsorption configurations for each pair. The green lines above and below the blue points indicate CatBERTa's intrinsic error threshold, with an average value of 0.24 eV for the exemplary twelve systems. This threshold for each adsorbate-catalyst combination is derived from the standard deviation of CatBERTa's predictions on the actual strings of those adsorption configurations, corresponding to the blue points. Since the CatBERTa prediction itself has intrinsic error from the DFT values, we add this value to the minimum and subtract it from the maximum values of the blue points, respectively, to establish the error threshold.

Figure \ref{fig:energy_plot} (b) demonstrates that incorporating LLM-derived configuration strings significantly increases the likelihood of CatBERTa’s predictions falling within the DFT energy and threshold ranges. After incorporating the LLM-derived configuration strings, the likelihood of CatBERTa’s predictions being within the target range more than doubles for both cases. This suggests that, despite the generated structures not being highly accurate, the information extracted from them can still benefit downstream prediction tasks.

\section{Conclusion}
We introduced a multimodal pretraining approach that integrates graph and text embeddings within the latent space. This approach facilitates connections between different model setups, enhancing the application of language models in prediction tasks. In the field of catalysis, the predictive language model CatBERTa can handle textual data with various features. Our graph-assisted pretraining method improves the accuracy of the language model by guiding the text modality using the graph modality. Additionally, we leveraged the autoregressive generative capabilities of the language model to predict energies without requiring precise atomic structures. By using a transformer-based language model to generate input strings, we create textual representations for the CatBERTa model based solely on chemical composition and surface orientation. This allows us to make energy predictions for adsorbate-catalyst systems using experimentally obtainable information. While the current framework has limitations in prediction accuracy and generation validity, it provides starting points for more detailed simulations or experimental validation.

Recent advancements in large language models (LLMs) have significantly impacted chemistry and material science, especially in areas like autoregressive generation, data retrieval, and autonomous scientific discovery\cite{coscientist, chemcrow, llm3dprint, llm_design}. Our study explores the potential of LLMs for both generative and predictive modeling. Moving forward, we aim to develop a more comprehensive, language-based platform for catalyst design by improving both predictive and generative capabilities, integrating them into a single LLM, incorporating additional functional tools, and equipping the platform with reasoning and planning capabilities in an agent-like framework.

\section{Methods}
\subsection{Open Catalyst dataset}

The OC20 dataset stands as the most extensive and varied dataset for heterogeneous catalysts. It encompasses over 1.2 million DFT relaxations, all using the revised Perdew-Burke-Emzerhof (RPBE) functional \cite{OC20, rpbe}. This dataset features various tasks, including Initial-Structure-to-Energy (IS2RE), Initial-Structure-to-Relaxed-Structure (IS2RS), and Structure-to-Energy-and-Force (S2EF). In our study, we focus on the data for the IS2RE/IS2RS task, which consists of 460,328 DFT relaxations. Our objective is to predict the relaxed energy of each adsorbate-catalyst configuration based on its final relaxed structure, leading us to specifically select the last frame of these relaxation trajectories.

To investigate the global minimum energy, also known as the adsorption energy, of adsorbate-catalyst pairs, the OC20-Dense dataset was developed \cite{adsorbml}. The OC20 dataset, while extensive in types of adsorbates and catalytic surfaces, lacks variation in adsorption configurations. OC20-Dense dataset addresses this by densely enumerating these configurations. The initial configurations of adsorbates on surfaces are produced using both heuristic and random approaches \cite{adsorbml, catkit, pymatgen}. These configurations then undergo relaxations using both ML and DFT methods. The OC20-Dense dataset contains 995 distinct adsorbate-catalyst pairs, evenly selected from the in-domain and three out-of-domain splits, from the OC20 validation set (see Figure \ref{fig:inference}). As a result, the entire OC20-Dense set has no overlap with the OC20 training set. The OC20-Dense training set, drawn from the entire OC20-Dense data set, comprises 15,450 data entries and serves as an optional addition. The validation set for the OC20-Dense is created by randomly selecting 9,001 data entries from the three out-of-domain splits of the OC20 validation set. For the test set, we used ML-relaxed structures from the OC20-Dense dataset, specifically from the same three out-of-domain splits in the OC20 validation set, provided by the Open Catalyst Challenge 2023 \cite{opencatalyst2023}.

\subsection{GNN-relaxed structures}

As part of the Open Catalyst Challenge 2023, the Open Catalyst Project has provided a set of ML-relaxed structures along with their energies calculated using DFT. These structures, originating from the OC20-Dense validation set and illustrated in Figure \ref{fig:inference} (b), were relaxed using models like GemNet-OC, SCN, and eSCN. Any relaxed structures that are invalid or lack valid DFT energy values were excluded by the dataset creator\cite{adsorbml, opencatalyst2023}. This includes cases where the adsorbate fails to bind to the surface, decomposes into different atoms or molecules, or causes significant alterations to the surface from its original state. After filtering out these invalid configurations, the remaining counts for the ML-relaxed test sets using the GemNet-OC, SCN, and eSCN models are 11,508, 11,630, and 11,755 structures, respectively. Within these datasets, only a subset of structures—919, 922, and 922 respectively—have valid DFT-verified energy values. Our accuracy analysis concentrates on these approximately 920 ML-relaxed structures, each supported by a reliable DFT energy assessment. Meanwhile, the embedding and attention score analyses fully utilize predictions on all the valid ML-relaxed structures.

\subsection{Structure-string conversion}
The input data is entirely text-based, adhering to the string-type input format outlined in the original CatBERTa paper \cite{catberta}. We generate textual strings by converting the relaxed structures in the OC20 and OC20-Dense datasets, as illustrated in Figure \ref{fig:inference}. Our textual input format is structured into three segments: the adsorbate, the catalytic surface, and the depiction of the adsorption configuration. Specifically, the adsorbate segment simply contains its elemental symbol. For the catalytic surface part, we integrate information about the catalyst's overall composition along with its Miller index. For these two segments, the information is sourced from the pre-existing metadata of the OC20 dataset. The depiction of the adsorption configuration is achieved by pinpointing both the primary and secondary atoms involved in the interaction. This method is selected due to its proven effectiveness in predicting energy outcomes in previous research \cite{catberta}. In this process, we identify these interacting elements using the Pymatgen library. First, we establish atomic connectivity based on a predefined cutoff radius, which is a covalent radius of the atom. Then, we pinpoint the atoms connected to those in the adsorbate and surface. The connected atoms of the adsorbate atoms are classified as primary interacting atoms, while the neighboring atoms of the primary interacting atoms on the surface are grouped as secondary interacting atoms.

To convert structures from LLM-generated CIFs, we employ a more lenient and simplified approach. Initially, we identify and specify only the adsorbate atom closest to the surface. Next, we gather the primary neighbor atoms surrounding this adsorbate atom. Following this, we collect the secondary neighbor atoms from the primary neighbors. In this process, we use a multiplier of 4 for the cut-off radius, meaning the neighbor atoms are those within four times the covalent radius. This approach is based on the understanding that the structures in the generated CIF are indicative, not exact.

\subsection{CatBERTa}
In this research, we employ the CatBERTa model as a predictive language model. This text-based model is specifically designed and trained for predicting relaxed energy in adsorbate-catalyst systems. The model incorporates the RoBERTa encoder, originally pretrained on an extensive natural language corpus that includes resources such as BookCorpus and English Wikipedia, cumulatively exceeding 160GB \cite{liu2019roberta}. The RoBERTa, diverging from the conventional BERT model \cite{devlin2019bert} which masks a fixed 15\% of tokens in each epoch during training, adopts a dynamic masking approach. This method alters the masked tokens variably across different training epochs, thereby improving the model’s proficiency in predicting masked words and grasping syntactic and semantic nuances.

The CatBERTa model is fine-tuned for an energy prediction task. The original RoBERTa’s classification head is replaced with a regression head, comprising a linear and activation layer. This modification allows CatBERTa to generate a singular scalar value of energy predictions. For this prediction, the embedding of the special ``\textless{}s\textgreater{}" token, after encoder processing, serves as input for this regression head. The training hyperparameters and the architecture details of the CatBERTa model are provided in Supplementary Table S1, while pretraining and fine-tuning strategies are listed in Supplementary Table S4.

\subsection{EquiformerV2}
The Equiformer is a GNN which is SE(3)/E(3)-equivariant, adeptly fusing the inductive biases of equivariance with the dynamic strengths of Transformers \cite{equiformer}. The Equiformer stands out by demonstrating that Transformers can be effectively adapted to 3D atomistic graphs. This is achieved by two main factors. First, the Equiformer modifies the traditional Transformer by substituting SE(3)/E(3)-equivariant operations for the original operations. Second, the Equiformer model introduces equivariant graph attention, a novel attention mechanism.

In the pretraining stage, we utilize the EquiformerV2 embedding for graph representation purposes due to its excellent performance in the OC20 dataset. The EquiformerV2 \cite{equiformerv2}, a refined version of the original Equiformer, brings to the table a host of enhancements. These improvements encompass the replacement of SO(3) convolutions with eSCN convolutions, the introduction of attention re-normalization, the incorporation of separable S2 activation, and the application of separable layer normalization\cite{equiformerv2}. Such advancements have elevated the EquiformerV2, especially in the task of energy and force predictions on the OC20 dataset. The model demonstrates state-of-the-art accuracy in its performance, achieving an MAE of 0.22 eV for the S2EF task and 0.31 eV for the IS2RE task, outperforming other benchmarked models. 

The graph embeddings in our case, are extracted after the final layer normalization in the EquiformerV2 model, which precedes the energy and force prediction stage, as illustrated in Figure \ref{fig:framework} (c). Within this model, each atom is represented as a node, and each atom node is characterized by a two-dimensional embedding tensor, collectively forming a three-dimensional tensor for the entire system. The size of the system tensor is defined by the number of atoms, the count of spherical channels, and the maximum degree of spherical harmonics involved \cite{equiformerv2}. The extraction of graph embeddings begins with reshaping the two-dimensional atom embedding tensor into a one-dimensional tensor. Subsequently, max-pooling is applied across all these one-dimensional atom embeddings in the system, yielding a single, comprehensive embedding for each system. In our study, the final embedding of each system is represented by a tensor with a size of 3,200. Consequently, during the graph-assisted pretraining phase, text embeddings, initially sized at 768, undergo a linear projection head to match the 3,200-size tensor.

\subsection{Contrastive loss}
Graph-assisted pretraining synergistically aligns text and graph embeddings through a self-supervised framework. Utilizing the graph encoder EquiformerV2 in a static (frozen) state, this method is specifically designed to transfer the insights from the graph to the text modality.

Inspired by the methodology used in Contrastive Language-Image Pretraining (CLIP) \cite{clip}, the graph-assisted pretraining strategy incorporates both text and graph encoders. The pretraining mechanism is centered around the optimization of a symmetric cross-entropy loss \cite{contrastive}. This optimization process aims to increase the similarity between embeddings from matching text-graph pairs while decreasing the similarity between those from unmatching pairs. By using a contrastive loss function, the primary objective is to establish a meaningful correlation between text and graph embeddings. The overarching goal is to precisely align embeddings from corresponding text-graph pairs while effectively differentiating between non-corresponding pairs.

The mathematical formulation of the loss function is defined as follows:

\begin{equation}
L = -\frac{1}{N} \sum_{i=1}^{N} \log \frac{e^{\text{sim}(I_i, T_i)/\tau}}{\sum_{j=1}^{N} e^{\text{sim}(I_i, T_j)/\tau} \mathbb{I}_{\{i \neq j\}}} - \frac{1}{N} \sum_{i=1}^{N} \log \frac{e^{\text{sim}(T_i, I_i)/\tau}}{\sum_{j=1}^{N} e^{\text{sim}(T_i, I_j)/\tau} \mathbb{I}_{\{i \neq j\}}}
\end{equation}

In this expression, \( T_i \) and \( G_i \) represent the embeddings of the i-th text and graph, respectively. The function \( \text{sim}(G_i, T_j) \) calculates the cosine similarity between the embeddings of the i-th graph and j-th text. Additionally, \( \tau \) is introduced as a temperature parameter, serving to appropriately scale the similarity scores within the model.

\subsection{Fine-tuning CrystaLLM}
CrystaLLM is a GPT-2-based large language model designed to generate crystal structures in CIF format for a given composition and, optionally, a specified space group \cite{crystallm}. Adapted from the nanoGPT\cite{nanoGPT} model implementation, the model is trained from scratch with a vocabulary size of 371, specifically for inorganic crystal systems. The training data was sourced from various databases, including the Materials Project \cite{mp}, the Open Quantum Materials Database \cite{oqmd}, and NOMAD \cite{nomad}. The tokenizer for the CrystaLLM operates on character bytes. The CrystaLLM (large) model, which we use in our framework, consists of 16 layers with 16 heads each, and individual block sizes of 2,048. This model has been trained for 48,000 epochs, enabling it to predict text-based structural CIF representations of various crystal systems by leveraging the generative capabilities of language models.

By fine-tuning the pretrained model over the combined training data from the OC20 and OC20-Dense datasets for 6,000 epochs, we leverage the learned embedding space of the existing model. This allows us to transfer the corresponding knowledge to our specific task. The fine-tuning learning rate is set to $6 \text{e-4}$, compared to the pretraining learning rate of $1 \text{e-3}$, to further enhance the model's generalization capabilities. Consequently, our fine-tuned CrystaLLM generates CIFs for target adsorbate-catalyst systems using only the adsorbate and catalyst bulk chemical symbols, along with surface orientation.

\subsection{CrystaLLM subsampling}
In this study, we applied targeted selection criteria to extract adsorbate-catalyst pairs from the OC20-Dense training dataset. Two filters were used to derive the final 66 downsampled adsorbate-catalyst pairs. The first filter ensures the number of atoms conforms to the token limits of the original CrystaLLM model, while the second filter focuses on composition matching to maintain a minimum level of accuracy in the generated configurations.

As we fine-tuned the original CrystaLLM model, we adhered to the token length constraints of the tokenizer, which is set at 3,000. We selected systems from the OC20-Dense dataset where the adsorbates contain no more than five atoms each, across 51 unique adsorbates. Notably, each adsorbate is composed of a maximum of three different types of elements, resulting in a total of 27 unique, filtered adsorbates. We also imposed a constraint on the number of atoms in the catalyst, selecting catalyst systems with no more than 72 atoms. Applying these filters to the OC20-Dense training dataset yielded 108 unique adsorbate-catalyst pairs out of the initial 235 pairs.

To ensure a minimum level of generation accuracy, we filtered out nonsensical compositions in the generated CIFs. Specifically, the adsorbate had to exactly match the given composition, while we allowed a tolerance of up to 12 atoms of deviation in the catalyst. By applying this composition-matching filter, the 108 adsorbate-catalyst pairs were reduced to 66. Due to the stochastic nature of the generative model, the number of valid systems can vary across iterations.

Following this process, we selected one CIF file for each pair that passed the criteria and converted it into the desired string configurations. If more than one generation out of three passed the criteria for a given pair, we randomly selected one CIF. We randomly selected 14 pairs from the in-domain split for exemplary visualization in Figure \ref{fig:energy_plot}. It is noteworthy that the 66 adsorbate-catalyst systems correspond to 5,141 overall configurations in the original training dataset, providing a total of 5,141 DFT-calculated energy values.

\begin{suppinfo}
Hyperparameters and architecture, Data split, Pretraining and Fine-tuning, Alignment in latent space, Results from diverse ML-relaxed structures, Prediction visualization, Prediction results of duplicate text sets
\end{suppinfo}

\section*{Technology Use Disclosure}
ChatGPT was used to help prepare the preprint version of this manuscript, specifically for grammar and typo corrections. All information in this manuscript has been read, corrected, and verified by all authors.

\section*{Data Availability Statement}
Access to the Open Catalyst 2020 dataset is provided via this link: \url{https://github.com/FAIR-Chem/fairchem}. The Open Catalyst 2020 Dense dataset and relevant data about the Open Catalyst Challenge 2023 are available at: \url{https://github.com/Open-Catalyst-Project/AdsorbML}. The preprocessed data, formatted for compatibility with the training framework, is available at the following link: \url{https://doi.org/10.6084/m9.figshare.27208356.v2}\cite{ock2024data}.

\section*{Code Availability Statement}
The Python code used in this study is available on Zenodo at \url{https://doi.org/10.5281/zenodo.13917199}\cite{ock2024code} and on GitHub at \url{https://github.com/hoon-ock/multi-view}.

\begin{acknowledgement}

We thank Meta Fundamental AI Research (FAIR) for providing the publicly available Open Catalyst Project dataset and organizing the Open Catalyst Challenge.

\end{acknowledgement}

\section*{Author declarations}
\subsection*{Conflict of Interest}
The authors have no conflicts to disclose.

\subsection*{Author Contributions}
J.O., S.B., and A.B.F. designed the research study. J.O., S.B., R.M., and A.A. developed the method, wrote the code, and performed the analysis. All authors wrote and approved the manuscript.

\bibliography{reference}

\providecommand{\latin}[1]{#1}
\makeatletter
\providecommand{\doi}
  {\begingroup\let\do\@makeother\dospecials
  \catcode`\{=1 \catcode`\}=2 \doi@aux}
\providecommand{\doi@aux}[1]{\endgroup\texttt{#1}}
\makeatother
\providecommand*\mcitethebibliography{\thebibliography}
\csname @ifundefined\endcsname{endmcitethebibliography}  {\let\endmcitethebibliography\endthebibliography}{}
\begin{mcitethebibliography}{45}
\providecommand*\natexlab[1]{#1}
\providecommand*\mciteSetBstSublistMode[1]{}
\providecommand*\mciteSetBstMaxWidthForm[2]{}
\providecommand*\mciteBstWouldAddEndPuncttrue
  {\def\EndOfBibitem{\unskip.}}
\providecommand*\mciteBstWouldAddEndPunctfalse
  {\let\EndOfBibitem\relax}
\providecommand*\mciteSetBstMidEndSepPunct[3]{}
\providecommand*\mciteSetBstSublistLabelBeginEnd[3]{}
\providecommand*\EndOfBibitem{}
\mciteSetBstSublistMode{f}
\mciteSetBstMaxWidthForm{subitem}{(\alph{mcitesubitemcount})}
\mciteSetBstSublistLabelBeginEnd
  {\mcitemaxwidthsubitemform\space}
  {\relax}
  {\relax}

\bibitem[Behler(2016)]{mlp}
Behler,~J. {Perspective: Machine learning potentials for atomistic simulations}. \emph{The Journal of Chemical Physics} \textbf{2016}, \emph{145}, 170901\relax
\mciteBstWouldAddEndPuncttrue
\mciteSetBstMidEndSepPunct{\mcitedefaultmidpunct}
{\mcitedefaultendpunct}{\mcitedefaultseppunct}\relax
\EndOfBibitem
\bibitem[Zitnick \latin{et~al.}(2020)Zitnick, Chanussot, Das, Goyal, Heras-Domingo, Ho, Hu, Lavril, Palizhati, Riviere, Shuaibi, Sriram, Tran, Wood, Yoon, Parikh, and Ulissi]{OC20_intro}
Zitnick,~C.~L. \latin{et~al.}  An Introduction to Electrocatalyst Design using Machine Learning for Renewable Energy Storage. 2020; \url{https://arxiv.org/abs/2010.09435}\relax
\mciteBstWouldAddEndPuncttrue
\mciteSetBstMidEndSepPunct{\mcitedefaultmidpunct}
{\mcitedefaultendpunct}{\mcitedefaultseppunct}\relax
\EndOfBibitem
\bibitem[Chanussot \latin{et~al.}(2021)Chanussot, Das, Goyal, Lavril, Shuaibi, Riviere, Tran, Heras-Domingo, Ho, Hu, Palizhati, Sriram, Wood, Yoon, Parikh, Zitnick, and Ulissi]{OC20}
Chanussot,~L. \latin{et~al.}  Open Catalyst 2020 (OC20) Dataset and Community Challenges. \emph{ACS Catalysis} \textbf{2021}, \emph{11}, 6059--6072\relax
\mciteBstWouldAddEndPuncttrue
\mciteSetBstMidEndSepPunct{\mcitedefaultmidpunct}
{\mcitedefaultendpunct}{\mcitedefaultseppunct}\relax
\EndOfBibitem
\bibitem[Reiser \latin{et~al.}(2022)Reiser, Neubert, Eberhard, Torresi, Zhou, Shao, Metni, van Hoesel, Schopmans, Sommer, and Friederich]{Reiser2022}
Reiser,~P.; Neubert,~M.; Eberhard,~A.; Torresi,~L.; Zhou,~C.; Shao,~C.; Metni,~H.; van Hoesel,~C.; Schopmans,~H.; Sommer,~T.; Friederich,~P. Graph neural networks for materials science and chemistry. \emph{Communications Materials} \textbf{2022}, \emph{3}, 93\relax
\mciteBstWouldAddEndPuncttrue
\mciteSetBstMidEndSepPunct{\mcitedefaultmidpunct}
{\mcitedefaultendpunct}{\mcitedefaultseppunct}\relax
\EndOfBibitem
\bibitem[Goldsmith \latin{et~al.}(2018)Goldsmith, Esterhuizen, Liu, Bartel, and Sutton]{MLforCat}
Goldsmith,~B.~R.; Esterhuizen,~J.; Liu,~J.-X.; Bartel,~C.~J.; Sutton,~C. Machine Learning for Heterogeneous Catalyst Design and Discovery. \emph{AIChE Journal} \textbf{2018}, \emph{64}, 2311--2323\relax
\mciteBstWouldAddEndPuncttrue
\mciteSetBstMidEndSepPunct{\mcitedefaultmidpunct}
{\mcitedefaultendpunct}{\mcitedefaultseppunct}\relax
\EndOfBibitem
\bibitem[Tran \latin{et~al.}(2022)Tran, Wang, Kingsbury, Palizhati, Persson, Jain, and Ulissi]{Tran2022}
Tran,~R.; Wang,~D.; Kingsbury,~R.; Palizhati,~A.; Persson,~K.~A.; Jain,~A.; Ulissi,~Z.~W. {Screening of bimetallic electrocatalysts for water purification with machine learning}. \emph{The Journal of Chemical Physics} \textbf{2022}, \emph{157}, 074102\relax
\mciteBstWouldAddEndPuncttrue
\mciteSetBstMidEndSepPunct{\mcitedefaultmidpunct}
{\mcitedefaultendpunct}{\mcitedefaultseppunct}\relax
\EndOfBibitem
\bibitem[Cao \latin{et~al.}(2024)Cao, Barati~Farimani, Ock, and Barati~Farimani]{ml_membrane}
Cao,~Z.; Barati~Farimani,~O.; Ock,~J.; Barati~Farimani,~A. Machine Learning in Membrane Design: From Property Prediction to AI-Guided Optimization. \emph{Nano Letters} \textbf{2024}, \emph{24}, 2953--2960\relax
\mciteBstWouldAddEndPuncttrue
\mciteSetBstMidEndSepPunct{\mcitedefaultmidpunct}
{\mcitedefaultendpunct}{\mcitedefaultseppunct}\relax
\EndOfBibitem
\bibitem[ope(2023)]{opencatalyst2023}
Open Catalyst Challenge. \url{https://opencatalystproject.org/challenge.html}, 2023; Accessed: December 24, 2023\relax
\mciteBstWouldAddEndPuncttrue
\mciteSetBstMidEndSepPunct{\mcitedefaultmidpunct}
{\mcitedefaultendpunct}{\mcitedefaultseppunct}\relax
\EndOfBibitem
\bibitem[Xie and Grossman(2018)Xie, and Grossman]{cgcnn}
Xie,~T.; Grossman,~J.~C. Crystal Graph Convolutional Neural Networks for an Accurate and Interpretable Prediction of Material Properties. \emph{Phys. Rev. Lett.} \textbf{2018}, \emph{120}, 145301\relax
\mciteBstWouldAddEndPuncttrue
\mciteSetBstMidEndSepPunct{\mcitedefaultmidpunct}
{\mcitedefaultendpunct}{\mcitedefaultseppunct}\relax
\EndOfBibitem
\bibitem[Schütt \latin{et~al.}(2017)Schütt, Kindermans, Sauceda, Chmiela, Tkatchenko, and Müller]{schnet}
Schütt,~K.~T.; Kindermans,~P.-J.; Sauceda,~H.~E.; Chmiela,~S.; Tkatchenko,~A.; Müller,~K.-R. SchNet: A continuous-filter convolutional neural network for modeling quantum interactions. 2017; \url{https://arxiv.org/abs/1706.08566}\relax
\mciteBstWouldAddEndPuncttrue
\mciteSetBstMidEndSepPunct{\mcitedefaultmidpunct}
{\mcitedefaultendpunct}{\mcitedefaultseppunct}\relax
\EndOfBibitem
\bibitem[Gasteiger \latin{et~al.}(2022)Gasteiger, Shuaibi, Sriram, Günnemann, Ulissi, Zitnick, and Das]{gemnet-oc}
Gasteiger,~J.; Shuaibi,~M.; Sriram,~A.; Günnemann,~S.; Ulissi,~Z.; Zitnick,~C.~L.; Das,~A. GemNet-OC: Developing Graph Neural Networks for Large and Diverse Molecular Simulation Datasets. 2022; \url{https://arxiv.org/abs/2204.02782}\relax
\mciteBstWouldAddEndPuncttrue
\mciteSetBstMidEndSepPunct{\mcitedefaultmidpunct}
{\mcitedefaultendpunct}{\mcitedefaultseppunct}\relax
\EndOfBibitem
\bibitem[Pablo-García \latin{et~al.}(2023)Pablo-García, Morandi, Vargas-Hernández, Jorner, Žarko Ivković, López, and Aspuru-Guzik]{reviewer3}
Pablo-García,~S.; Morandi,~S.; Vargas-Hernández,~R.~A.; Jorner,~K.; Žarko Ivković; López,~N.; Aspuru-Guzik,~A. Fast evaluation of the adsorption energy of organic molecules on metals via graph neural networks. \emph{Nature Computational Science} \textbf{2023}, \emph{3}, 433--442\relax
\mciteBstWouldAddEndPuncttrue
\mciteSetBstMidEndSepPunct{\mcitedefaultmidpunct}
{\mcitedefaultendpunct}{\mcitedefaultseppunct}\relax
\EndOfBibitem
\bibitem[Studt(2021)]{challenge}
Studt,~F. Grand Challenges in Computational Catalysis. \emph{Frontiers in Catalysis} \textbf{2021}, \emph{1}\relax
\mciteBstWouldAddEndPuncttrue
\mciteSetBstMidEndSepPunct{\mcitedefaultmidpunct}
{\mcitedefaultendpunct}{\mcitedefaultseppunct}\relax
\EndOfBibitem
\bibitem[Giulimondi \latin{et~al.}(2023)Giulimondi, Mitchell, and Pérez-Ramírez]{challenge2}
Giulimondi,~V.; Mitchell,~S.; Pérez-Ramírez,~J. Challenges and Opportunities in Engineering the Electronic Structure of Single-Atom Catalysts. \emph{ACS Catalysis} \textbf{2023}, \emph{13}, 2981--2997\relax
\mciteBstWouldAddEndPuncttrue
\mciteSetBstMidEndSepPunct{\mcitedefaultmidpunct}
{\mcitedefaultendpunct}{\mcitedefaultseppunct}\relax
\EndOfBibitem
\bibitem[Cao \latin{et~al.}(2023)Cao, Magar, Wang, and Barati~Farimani]{moformer}
Cao,~Z.; Magar,~R.; Wang,~Y.; Barati~Farimani,~A. MOFormer: Self-Supervised Transformer Model for Metal–Organic Framework Property Prediction. \emph{Journal of the American Chemical Society} \textbf{2023}, \emph{145}, 2958--2967\relax
\mciteBstWouldAddEndPuncttrue
\mciteSetBstMidEndSepPunct{\mcitedefaultmidpunct}
{\mcitedefaultendpunct}{\mcitedefaultseppunct}\relax
\EndOfBibitem
\bibitem[Balaji and Magar(2023)Balaji, and Magar]{gpt-molberta}
Balaji,~S.; Magar,~R. GPT-MolBERTa: GPT Molecular Features Language Model for molecular property prediction. 2023; \url{https://arxiv.org/abs/2310.03030}\relax
\mciteBstWouldAddEndPuncttrue
\mciteSetBstMidEndSepPunct{\mcitedefaultmidpunct}
{\mcitedefaultendpunct}{\mcitedefaultseppunct}\relax
\EndOfBibitem
\bibitem[Xu \latin{et~al.}(2023)Xu, Wang, and Barati~Farimani]{transpolymer}
Xu,~C.; Wang,~Y.; Barati~Farimani,~A. TransPolymer: a Transformer-based language model for polymer property predictions. \emph{npj Computational Materials} \textbf{2023}, \emph{9}, 64\relax
\mciteBstWouldAddEndPuncttrue
\mciteSetBstMidEndSepPunct{\mcitedefaultmidpunct}
{\mcitedefaultendpunct}{\mcitedefaultseppunct}\relax
\EndOfBibitem
\bibitem[Ock \latin{et~al.}(2023)Ock, Guntuboina, and Barati~Farimani]{catberta}
Ock,~J.; Guntuboina,~C.; Barati~Farimani,~A. Catalyst Energy Prediction with CatBERTa: Unveiling Feature Exploration Strategies through Large Language Models. \emph{ACS Catalysis} \textbf{2023}, \emph{13}, 16032--16044\relax
\mciteBstWouldAddEndPuncttrue
\mciteSetBstMidEndSepPunct{\mcitedefaultmidpunct}
{\mcitedefaultendpunct}{\mcitedefaultseppunct}\relax
\EndOfBibitem
\bibitem[Wang \latin{et~al.}(2011)Wang, Temel, Shen, Jones, Grabow, Studt, Bligaard, Abild‐Pedersen, Christensen, and N{\o}rskov]{BEP}
Wang,~S.; Temel,~B.; Shen,~J.; Jones,~G.; Grabow,~L.~C.; Studt,~F.; Bligaard,~T.; Abild‐Pedersen,~F.; Christensen,~C.~H.; N{\o}rskov,~J.~K. Universal Br{\o}nsted-Evans-Polanyi Relations for C–C, C–O, C–N, N–O, N–N, and O–O Dissociation Reactions. \emph{Catalysis Letters} \textbf{2011}, \emph{141}, 370--373\relax
\mciteBstWouldAddEndPuncttrue
\mciteSetBstMidEndSepPunct{\mcitedefaultmidpunct}
{\mcitedefaultendpunct}{\mcitedefaultseppunct}\relax
\EndOfBibitem
\bibitem[Sutton and Vlachos(2012)Sutton, and Vlachos]{BEP2}
Sutton,~J.~E.; Vlachos,~D.~G. A Theoretical and Computational Analysis of Linear Free Energy Relations for the Estimation of Activation Energies. \emph{ACS Catalysis} \textbf{2012}, \emph{2}, 1624--1634\relax
\mciteBstWouldAddEndPuncttrue
\mciteSetBstMidEndSepPunct{\mcitedefaultmidpunct}
{\mcitedefaultendpunct}{\mcitedefaultseppunct}\relax
\EndOfBibitem
\bibitem[Ock \latin{et~al.}(2023)Ock, Tian, Kitchin, and Ulissi]{Ock2023}
Ock,~J.; Tian,~T.; Kitchin,~J.; Ulissi,~Z. {Beyond independent error assumptions in large GNN atomistic models}. \emph{The Journal of Chemical Physics} \textbf{2023}, \emph{158}, 214702\relax
\mciteBstWouldAddEndPuncttrue
\mciteSetBstMidEndSepPunct{\mcitedefaultmidpunct}
{\mcitedefaultendpunct}{\mcitedefaultseppunct}\relax
\EndOfBibitem
\bibitem[Huang and Barati~Farimani(2024)Huang, and Barati~Farimani]{Huang2024}
Huang,~H.; Barati~Farimani,~A. {Multimodal learning of heat capacity based on transformers and crystallography pretraining}. \emph{Journal of Applied Physics} \textbf{2024}, \emph{135}, 165104\relax
\mciteBstWouldAddEndPuncttrue
\mciteSetBstMidEndSepPunct{\mcitedefaultmidpunct}
{\mcitedefaultendpunct}{\mcitedefaultseppunct}\relax
\EndOfBibitem
\bibitem[Badrinarayanan \latin{et~al.}(2024)Badrinarayanan, Guntuboina, Mollaei, and Farimani]{multipeptide}
Badrinarayanan,~S.; Guntuboina,~C.; Mollaei,~P.; Farimani,~A.~B. Multi-Peptide: Multimodality Leveraged Language-Graph Learning of Peptide Properties. 2024; \url{https://arxiv.org/abs/2407.03380}\relax
\mciteBstWouldAddEndPuncttrue
\mciteSetBstMidEndSepPunct{\mcitedefaultmidpunct}
{\mcitedefaultendpunct}{\mcitedefaultseppunct}\relax
\EndOfBibitem
\bibitem[Antunes \latin{et~al.}(2024)Antunes, Butler, and Grau-Crespo]{crystallm}
Antunes,~L.~M.; Butler,~K.~T.; Grau-Crespo,~R. Crystal Structure Generation with Autoregressive Large Language Modeling. 2024; \url{https://arxiv.org/abs/2307.04340}\relax
\mciteBstWouldAddEndPuncttrue
\mciteSetBstMidEndSepPunct{\mcitedefaultmidpunct}
{\mcitedefaultendpunct}{\mcitedefaultseppunct}\relax
\EndOfBibitem
\bibitem[Gruver \latin{et~al.}(2024)Gruver, Sriram, Madotto, Wilson, Zitnick, and Ulissi]{gruver2024meta}
Gruver,~N.; Sriram,~A.; Madotto,~A.; Wilson,~A.~G.; Zitnick,~C.~L.; Ulissi,~Z. Fine-Tuned Language Models Generate Stable Inorganic Materials as Text. 2024; \url{https://arxiv.org/abs/2402.04379}\relax
\mciteBstWouldAddEndPuncttrue
\mciteSetBstMidEndSepPunct{\mcitedefaultmidpunct}
{\mcitedefaultendpunct}{\mcitedefaultseppunct}\relax
\EndOfBibitem
\bibitem[Liao \latin{et~al.}(2023)Liao, Wood, Das, and Smidt]{equiformerv2}
Liao,~Y.-L.; Wood,~B.; Das,~A.; Smidt,~T. EquiformerV2: Improved Equivariant Transformer for Scaling to Higher-Degree Representations. 2023; \url{https://arxiv.org/abs/2306.12059}\relax
\mciteBstWouldAddEndPuncttrue
\mciteSetBstMidEndSepPunct{\mcitedefaultmidpunct}
{\mcitedefaultendpunct}{\mcitedefaultseppunct}\relax
\EndOfBibitem
\bibitem[Liao and Smidt(2023)Liao, and Smidt]{equiformer}
Liao,~Y.-L.; Smidt,~T. Equiformer: Equivariant Graph Attention Transformer for 3D Atomistic Graphs. 2023; \url{https://arxiv.org/abs/2206.11990}\relax
\mciteBstWouldAddEndPuncttrue
\mciteSetBstMidEndSepPunct{\mcitedefaultmidpunct}
{\mcitedefaultendpunct}{\mcitedefaultseppunct}\relax
\EndOfBibitem
\bibitem[Gao \latin{et~al.}(2020)Gao, Chen, Li, Liu, Liu, and Jiang]{Gao2020}
Gao,~W.; Chen,~Y.; Li,~B.; Liu,~S.-P.; Liu,~X.; Jiang,~Q. Determining the adsorption energies of small molecules with the intrinsic properties of adsorbates and substrates. \emph{Nature Communications} \textbf{2020}, \emph{11}, 1196\relax
\mciteBstWouldAddEndPuncttrue
\mciteSetBstMidEndSepPunct{\mcitedefaultmidpunct}
{\mcitedefaultendpunct}{\mcitedefaultseppunct}\relax
\EndOfBibitem
\bibitem[Esterhuizen \latin{et~al.}(2020)Esterhuizen, Goldsmith, and Linic]{Jacques2020}
Esterhuizen,~J.~A.; Goldsmith,~B.~R.; Linic,~S. Theory-Guided Machine Learning Finds Geometric Structure-Property Relationships for Chemisorption on Subsurface Alloys. \emph{Chem} \textbf{2020}, \emph{6}, 3100--3117\relax
\mciteBstWouldAddEndPuncttrue
\mciteSetBstMidEndSepPunct{\mcitedefaultmidpunct}
{\mcitedefaultendpunct}{\mcitedefaultseppunct}\relax
\EndOfBibitem
\bibitem[Boiko \latin{et~al.}(2023)Boiko, MacKnight, Kline, and Gomes]{coscientist}
Boiko,~D.~A.; MacKnight,~R.; Kline,~B.; Gomes,~G. Autonomous chemical research with large language models. \emph{Nature} \textbf{2023}, \emph{624}, 570--578\relax
\mciteBstWouldAddEndPuncttrue
\mciteSetBstMidEndSepPunct{\mcitedefaultmidpunct}
{\mcitedefaultendpunct}{\mcitedefaultseppunct}\relax
\EndOfBibitem
\bibitem[M.~Bran \latin{et~al.}(2024)M.~Bran, Cox, Schilter, Baldassari, White, and Schwaller]{chemcrow}
M.~Bran,~A.; Cox,~S.; Schilter,~O.; Baldassari,~C.; White,~A.~D.; Schwaller,~P. Augmenting large language models with chemistry tools. \emph{Nature Machine Intelligence} \textbf{2024}, 1--11\relax
\mciteBstWouldAddEndPuncttrue
\mciteSetBstMidEndSepPunct{\mcitedefaultmidpunct}
{\mcitedefaultendpunct}{\mcitedefaultseppunct}\relax
\EndOfBibitem
\bibitem[Jadhav \latin{et~al.}(2024)Jadhav, Pak, and Farimani]{llm3dprint}
Jadhav,~Y.; Pak,~P.; Farimani,~A.~B. LLM-3D Print: Large Language Models To Monitor and Control 3D Printing. 2024; \url{https://arxiv.org/abs/2408.14307}\relax
\mciteBstWouldAddEndPuncttrue
\mciteSetBstMidEndSepPunct{\mcitedefaultmidpunct}
{\mcitedefaultendpunct}{\mcitedefaultseppunct}\relax
\EndOfBibitem
\bibitem[Jadhav and Farimani(2024)Jadhav, and Farimani]{llm_design}
Jadhav,~Y.; Farimani,~A.~B. Large Language Model Agent as a Mechanical Designer. 2024; \url{https://arxiv.org/abs/2404.17525}\relax
\mciteBstWouldAddEndPuncttrue
\mciteSetBstMidEndSepPunct{\mcitedefaultmidpunct}
{\mcitedefaultendpunct}{\mcitedefaultseppunct}\relax
\EndOfBibitem
\bibitem[Hammer \latin{et~al.}(1999)Hammer, Hansen, and N\o{}rskov]{rpbe}
Hammer,~B.; Hansen,~L.~B.; N\o{}rskov,~J.~K. Improved Adsorption Energetics within Density-Functional Theory using Revised Perdew-Burke-Ernzerhof Functionals. \emph{Phys. Rev. B} \textbf{1999}, \emph{59}, 7413--7421\relax
\mciteBstWouldAddEndPuncttrue
\mciteSetBstMidEndSepPunct{\mcitedefaultmidpunct}
{\mcitedefaultendpunct}{\mcitedefaultseppunct}\relax
\EndOfBibitem
\bibitem[Liu \latin{et~al.}(2019)Liu, Ott, Goyal, Du, Joshi, Chen, Levy, Lewis, Zettlemoyer, and Stoyanov]{liu2019roberta}
Liu,~Y.; Ott,~M.; Goyal,~N.; Du,~J.; Joshi,~M.; Chen,~D.; Levy,~O.; Lewis,~M.; Zettlemoyer,~L.; Stoyanov,~V. RoBERTa: A Robustly Optimized BERT Pretraining Approach. 2019; \url{https://arxiv.org/abs/1907.11692}\relax
\mciteBstWouldAddEndPuncttrue
\mciteSetBstMidEndSepPunct{\mcitedefaultmidpunct}
{\mcitedefaultendpunct}{\mcitedefaultseppunct}\relax
\EndOfBibitem
\bibitem[Devlin \latin{et~al.}(2019)Devlin, Chang, Lee, and Toutanova]{devlin2019bert}
Devlin,~J.; Chang,~M.-W.; Lee,~K.; Toutanova,~K. BERT: Pre-training of Deep Bidirectional Transformers for Language Understanding. 2019; \url{https://arxiv.org/abs/1810.04805}\relax
\mciteBstWouldAddEndPuncttrue
\mciteSetBstMidEndSepPunct{\mcitedefaultmidpunct}
{\mcitedefaultendpunct}{\mcitedefaultseppunct}\relax
\EndOfBibitem
\bibitem[Radford \latin{et~al.}(2021)Radford, Kim, Hallacy, Ramesh, Goh, Agarwal, Sastry, Askell, Mishkin, Clark, Krueger, and Sutskever]{clip}
Radford,~A.; Kim,~J.~W.; Hallacy,~C.; Ramesh,~A.; Goh,~G.; Agarwal,~S.; Sastry,~G.; Askell,~A.; Mishkin,~P.; Clark,~J.; Krueger,~G.; Sutskever,~I. Learning Transferable Visual Models From Natural Language Supervision. 2021; \url{https://arxiv.org/abs/2103.00020}\relax
\mciteBstWouldAddEndPuncttrue
\mciteSetBstMidEndSepPunct{\mcitedefaultmidpunct}
{\mcitedefaultendpunct}{\mcitedefaultseppunct}\relax
\EndOfBibitem
\bibitem[van~den Oord \latin{et~al.}(2019)van~den Oord, Li, and Vinyals]{contrastive}
van~den Oord,~A.; Li,~Y.; Vinyals,~O. Representation Learning with Contrastive Predictive Coding. 2019; \url{https://arxiv.org/abs/1807.03748}\relax
\mciteBstWouldAddEndPuncttrue
\mciteSetBstMidEndSepPunct{\mcitedefaultmidpunct}
{\mcitedefaultendpunct}{\mcitedefaultseppunct}\relax
\EndOfBibitem
\bibitem[Karpathy(2022)]{nanoGPT}
Karpathy,~A. \text{NanoGPT}. \url{https://github.com/karpathy/nanoGPT}, 2022; Accessed: Aug 1, 2024\relax
\mciteBstWouldAddEndPuncttrue
\mciteSetBstMidEndSepPunct{\mcitedefaultmidpunct}
{\mcitedefaultendpunct}{\mcitedefaultseppunct}\relax
\EndOfBibitem
\bibitem[Jain \latin{et~al.}(2013)Jain, Ong, Hautier, Chen, Richards, Dacek, Cholia, Gunter, Skinner, Ceder, and Persson]{mp}
Jain,~A.; Ong,~S.~P.; Hautier,~G.; Chen,~W.; Richards,~W.~D.; Dacek,~S.; Cholia,~S.; Gunter,~D.; Skinner,~D.; Ceder,~G.; Persson,~K.~A. {Commentary: The Materials Project: A materials genome approach to accelerating materials innovation}. \emph{APL Materials} \textbf{2013}, \emph{1}, 011002\relax
\mciteBstWouldAddEndPuncttrue
\mciteSetBstMidEndSepPunct{\mcitedefaultmidpunct}
{\mcitedefaultendpunct}{\mcitedefaultseppunct}\relax
\EndOfBibitem
\bibitem[Saal \latin{et~al.}(2013)Saal, Kirklin, Aykol, Meredig, and Wolverton]{oqmd}
Saal,~J.~E.; Kirklin,~S.; Aykol,~M.; Meredig,~B.; Wolverton,~C. Materials Design and Discovery with High-Throughput Density Functional Theory: The Open Quantum Materials Database (OQMD). \emph{JOM} \textbf{2013}, \emph{65}, 1501--1509\relax
\mciteBstWouldAddEndPuncttrue
\mciteSetBstMidEndSepPunct{\mcitedefaultmidpunct}
{\mcitedefaultendpunct}{\mcitedefaultseppunct}\relax
\EndOfBibitem
\bibitem[nom(2023)]{nomad}
NOMAD: A distributed web-based platform for managing materials science research data. \emph{Journal of Open Source Software} \textbf{2023}, \emph{8}, 5388\relax
\mciteBstWouldAddEndPuncttrue
\mciteSetBstMidEndSepPunct{\mcitedefaultmidpunct}
{\mcitedefaultendpunct}{\mcitedefaultseppunct}\relax
\EndOfBibitem
\bibitem[Ock \latin{et~al.}(2024)Ock, Badrinarayanan, Magar, Antony, and Barati~Farimani]{ock2024data}
Ock,~J.; Badrinarayanan,~S.; Magar,~R.; Antony,~A.; Barati~Farimani,~A. Language and graph multimodal data for heterogeneous catalyst. 2024; \url{https://doi.org/10.6084/m9.figshare.27208356.v2}\relax
\mciteBstWouldAddEndPuncttrue
\mciteSetBstMidEndSepPunct{\mcitedefaultmidpunct}
{\mcitedefaultendpunct}{\mcitedefaultseppunct}\relax
\EndOfBibitem
\bibitem[Ock(2024)]{ock2024code}
Ock,~J. hoon-ock/multi-view: release. 2024; \url{https://doi.org/10.5281/zenodo.13922448}\relax
\mciteBstWouldAddEndPuncttrue
\mciteSetBstMidEndSepPunct{\mcitedefaultmidpunct}
{\mcitedefaultendpunct}{\mcitedefaultseppunct}\relax
\EndOfBibitem
\end{mcitethebibliography}

\setcounter{table}{0}
\renewcommand{\thetable}{S\arabic{table}}
\setcounter{figure}{0}
\renewcommand{\thefigure}{S\arabic{figure}}
\setcounter{equation}{0}
\renewcommand{\theequation}{S\arabic{equation}}
\newpage
\maketitle
\section{Supplementary Information}
\section{Hyperparameters and architecture}
\label{si:hyperparameter}

The CatBERTa encoder retains the same architecture of the publicly available RoBERTa encoder. Identical configurations, except the loss function, are employed in graph-assisted pretraining and the fine-tuning steps.

\begin{table}[htbp]
	\centering
	\caption{Overview of CatBERTa's architecture and training hyperparameters.}
	\label{tab:hyperparams}
	\begin{tabular}{lc}
	\hline
	Hyperparameter & Value \\
	\hline
    Max positional embeddings & 514 \\
	Number of attention heads & 12 \\
	Number of hidden layers & 12 \\
	Size of each hidden layer & 768 \\
    Dropout probability in hidden layer & 0.1\\
    Batch size & 32 \\
	Optimizer & AdamW \\
    Scheduler & reduceLR \\
	Initial learning rate & $1 \times 10^{-5}$ \\
	Early stopping threshold & 5 \\
	Warmup steps & 0 \\
	Loss function & MSE \\
	\hline
	\end{tabular}
\end{table}

The finetuning of the CrystaLLM model was performed using a refined set of hyperparameters that were optimized for transfer learning from the pretrained base model. The key adjustments made during finetuning included modifications to the learning rate, batch size, and gradient accumulation steps to account for smaller, domain-specific datasets and to balance the computational load. The model was initialized from the pretrained checkpoint, allowing the learning process to build upon the knowledge acquired during pretraining. No dropout was applied during this phase, as it was not required to prevent overfitting in the context of this specific fine-tuning task.

\begin{table}[h]
\centering
\caption{Overview of CrystaLLM's architecture and fine-tuning hyperparameters.}
\begin{tabular}{lc}
\hline
Hyperparameter                  & Value \\ \hline
Eval Interval                   & 20                            \\ 
Eval Iterations         	    & 200                          \\ 
Gradient Accumulation Steps     & 40                            \\ 
Batch Size                      & 2                             \\ 
Block Size                      & 2,048                         \\ 
Number of Layers                & 12                            \\ 
Number of Heads                 & 12                            \\ 
Embedding Size                  & 768                           \\ 
Dropout                         & 0.0                           \\ 
Learning Rate                   & 6e-4                          \\ 
Weight Decay                    & 0.1                           \\ 
\(\beta_1\) (AdamW)             & 0.9                           \\ 
\(\beta_2\) (AdamW)             & 0.95                          \\ 
Max Iterations                  & 6,000                         \\ 
Learning Rate Decay Iters       & 3,000                         \\ 
Minimum Learning Rate           & 6e-3                          \\ 
Warmup Iterations               & 200                           \\ 
Gradient Clipping               & 1.0                           \\ \hline
\end{tabular}
\end{table}

\newpage
\section{Data split}

Two types of training were conducted for the CatBERTa model: graph-assisted pretraining and downstream regression training using energy labels. The input textual strings were sourced from the Open Catalyst dataset, with 460k strings obtained from the relaxed structures in the OC20 IS2RE training set and 15.5k strings sourced from the OC20-Dense dataset. The OC20-Dense dataset was optionally used for graph-assisted pretraining and energy regression training. The OC20-Dense set was constructed by enumerating adsorption configurations over the OC20 validation set. Importantly, there is no overlap between the OC20 training set and the OC20-Dense dataset. The test was conducted on ML-relaxed structures from the OC20-Dense set, sourced from the three out-of-domain splits in the OC20 dataset, rather than on DFT-relaxed structures.

For CrystaLLM fine-tuning, we used both the 460k OC20 training set and the 15.5k OC20-Dense training set. Inference was performed autoregressively, predicting the next token based on an initial input prompt describing the adsorbate-catalyst pair, such as `data\_CHOHTe16Au8 (1 0 0)’. This means that for inference, only the chemical symbols and surface orientation of the adsorbate-catalyst pair were required. These pairs were collected from the OC20-Dense training set, and inferences were run to generate the corresponding CIFs. Although CIF generation was based on adsorbate-catalyst pairs from the OC20-Dense training set, subsequent CatBERTa predictions were made using a model trained solely on the OC20 training set, which has no overlap with the OC20-Dense training set. This ensures that the energy predictions remained fully out-of-domain.

\begin{table}[ht]
\centering
\caption{Data split for model training and test.}
\resizebox{\textwidth}{!}{%
\begin{tabular}{l p{6cm} p{6cm}}  
\toprule
               & CatBERTa & CrystaLLM \\ \midrule
Train      & 460k OC20 train set \newline \vspace{2mm} (\& 15.5k OC20-Dense train set) & 460k OC20 train set \newline \vspace{2mm} \& 15.5k OC20-Dense train set \\
Validation & 9k OC20-Dense val set & - \\
Test       & 11.5-11.8k ML-relaxed structures \newline \vspace{2mm} (919-922 w/ valid DFT energy) & 66 adsorbate-catalyst pairs \\ 
\bottomrule
\end{tabular}
}
\end{table}

\newpage
\section{Pretraining and Fine-tuning}

To evaluate the enhancement from graph-assisted pretraining, we compare the prediction results of CatBERTa with and without this pretraining method, as shown in Table \ref{tab:pretrain}. The results are benchmarked using the OC20 and OC20-Dense datasets. The CatBERTa model uses a RoBERTa encoder, which is initially pretrained on a large English corpus using masked language modeling for natural language processing tasks. Therefore, when we do not apply graph-assisted pretraining, we directly use RoBERTa, pretrained for natural language processing tasks, as the encoder. In this context, when not applying graph-assisted pretraining, this natural language modeling serves as the pretraining phase. For the energy prediction task, the model is fine-tuned using relaxed energy values as regression labels.

\begin{table}[htbp]
\centering
\caption{Overview of pretraining and fine-tuning. Graph-assisted pretraining and fine-tuning for energy modeling involves DFT-relaxed structure data sourced from the OC20 and OC20-Dense datasets. The term `Combined' refers to the combination of OC20 and OC20-Dense datasets, and `GAP' stands for graph-assisted pretraining.}
\label{tab:pretrain}
\resizebox{\textwidth}{!}{%
\begin{tabular}{lcccc}
\toprule
 & \multicolumn{2}{c}{Pretraining} & \multicolumn{2}{c}{Fine-tuning} \\
\cmidrule(lr){2-3} 
\cmidrule(lr){4-5}
 &  Method & Data Source & Method & Data Source\\
\midrule
CatBERTa & Masked language modeling & English corpus & Energy prediction & Open Catalyst Dataset\\
GAP-CatBERTa & Graph-assisted pretraining & Open Catalyst Dataset & Energy prediction & Open Catalyst Dataset \\

\bottomrule
\end{tabular}
}
\end{table}

\newpage
\section{Alignment in latent space}

\begin{figure*}[h!] 
\centering
\includegraphics[width=0.9\textwidth]{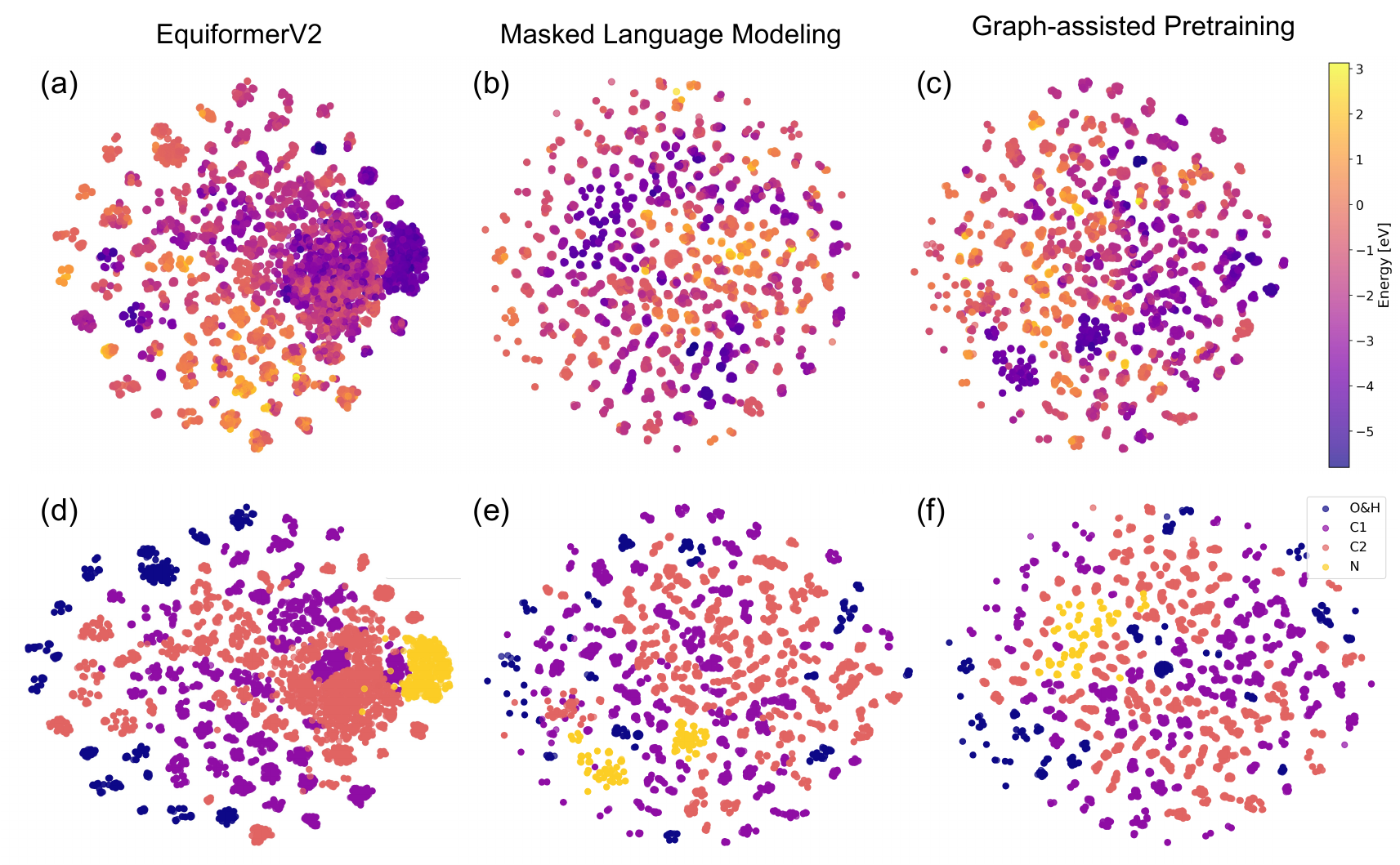} 
\caption{t-SNE visualizations of graph and text embeddings. Graph embeddings are derived from EquiformerV2, while text embeddings are from CatBERTa's ``\textless{}s\textgreater{}" token. Upper row panels \textbf{a-c} use colors to indicate energy levels, while lower row panels \textbf{d-f} color-code by adsorbate type. These adsorbate types are represented as follows: O\&H (oxygen/hydrogen), C1 (single carbon-containing adsorbates), C2 (two carbon-containing adsorbates), and N (nitrogen-containing adsorbates)}
\label{fig:tsne}
\end{figure*}

Our findings demonstrate that graph-assisted pretraining is relatively more effective in clustering systems in the latent space according to their energy levels and adsorbate types, compared to the encoder pretrained with masked language modeling. In Figure \ref{fig:tsne}, the EquiformerV2 embeddings (panels a and d) exhibit distinct clustering for systems with low energy and identical adsorbate types, exemplifying an ideal latent space organization. We presume this clustering precision is associated with the high accuracy of EquiformerV2 in energy predictions. Conversely, the encoder pretrained with masked language modeling results in a more scattered distribution of embeddings (panels b and e), though some degree of clustering exists. Systems with lower energy and nitrogen-containing adsorbates show more pronounced clustering in graph-assisted pretraining, compared to the masked language modeling approach (comparing panels b and c, e and f, respectively). Consequently, graph-assisted pretraining restructures the latent space to reflect the high-performance traits of EquiformerV2, thereby providing a more effective starting point for the subsequent fine-tuning process. 


\newpage
\section{Results from diverse ML-relaxed structures}

\begin{table}[h!]
\centering
\caption{CatBERTa prediction results from various ML-relaxed structures as generated by GemNet-OC, SCN, and eSCN.}
\label{tab:results}
\resizebox{\textwidth}{!}{%
\begin{tabular}{lcccccc}
\toprule
 & \multicolumn{3}{c}{MAE [eV] (\(\downarrow\))} & \multicolumn{3}{c}{\( R^2 \) (\(\uparrow\))} \\
\cmidrule(lr){2-4} 
\cmidrule(lr){5-7}
& GemNet-OC & SCN & eSCN & GemNet-OC  & SCN  & eSCN  \\
\hline
CatBERTa\textsubscript{OC20} & 0.696 & 0.720 & 0.722 & 0.597 & 0.584 & 0.570  \\ 
GAP-CatBERTa\textsubscript{OC20} & 0.620 & 0.654 & 0.656 & 0.708 & 0.687 & 0.679 \\ 
CatBERTa\textsubscript{combined} & 0.373 & 0.382 & 0.379 & 0.868 & 0.862 & 0.858  \\ 
GAP-CatBERTa\textsubscript{combined} & 0.345 & 0.351 & 0.341 & 0.882 & 0.880 & 0.883  \\
\bottomrule
\end{tabular}
}
\end{table}

\section{Prediction visualization}

\begin{figure*}[hbpt] 
\centering
\includegraphics[width=0.7\textwidth]{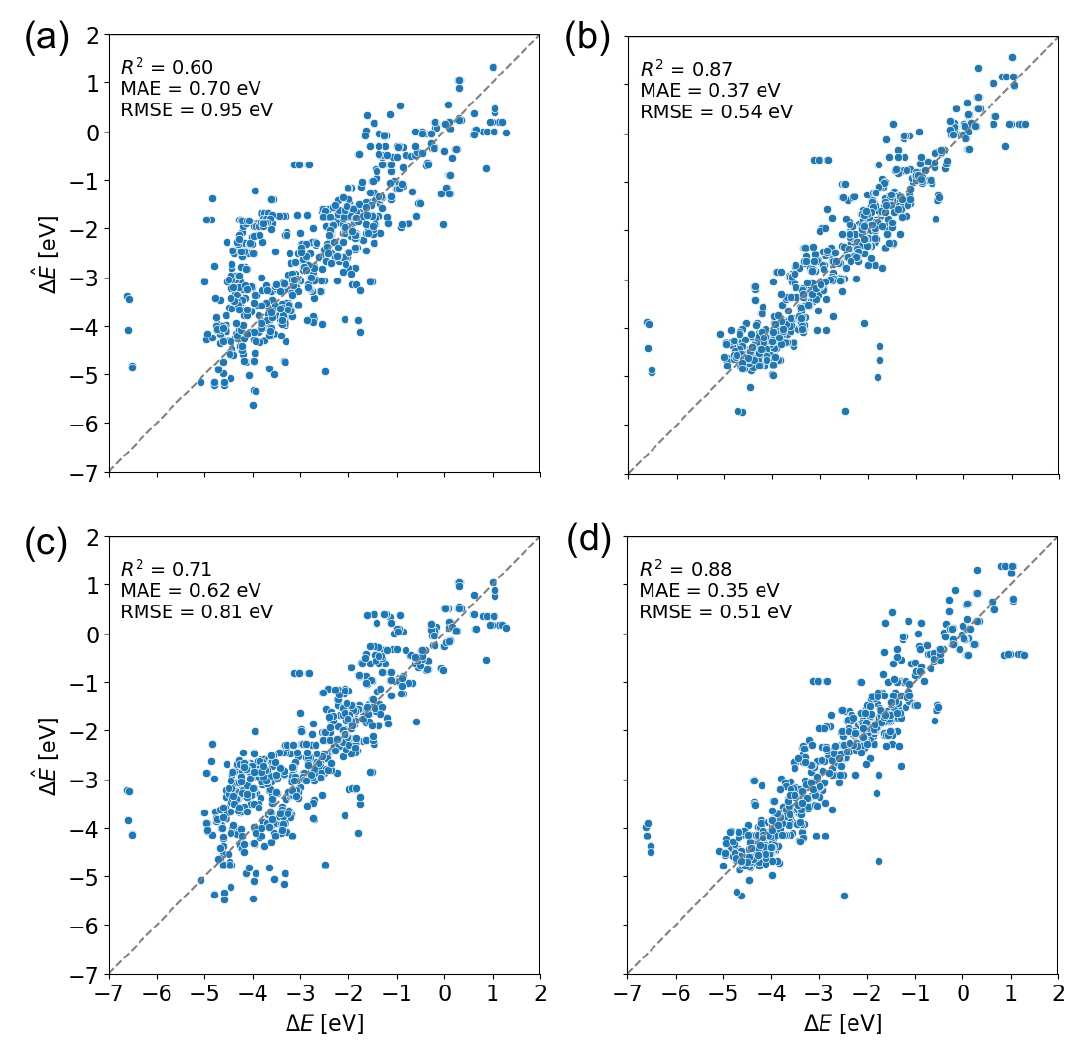} 
\caption{Parity plots comparing DFT-calculated (x-axis) and CatBERTa-predicted (y-axis) energy values for structures relaxed using GemNet-OC. All subfigures are aligned with identical x and y-axes for direct comparison. \textbf{a} and \textbf{c} show results without graph-assisted pretraining, while \textbf{b} and \textbf{d} present the results with it.}
\label{fig:parity}
\end{figure*}

The predictions made on GemNet-OC-relaxed structures were selected for example visualization, as shown in Figure \ref{fig:parity}. The trend follows the results shown in Table 1. The outlier clusters, located at a DFT energy of around -6.5 eV, are adsorption configurations of \ce{CHCH} adsorbate and \ce{Ti36Re18Os18} (0, 0, 1) catalyst. This is likely due to the poor prediction accuracy for the adsorbate \ce{CHCH}. All entries containing \ce{CHCH} show an MAE of 0.79 eV for the GAP-CatBERTa\textsubscript{combined} case, which is higher than the MAE for the entire dataset. Although their configurations differ, their energies are very similar, ranging from -6.60 to -6.51 eV, a difference of only 0.12 eV. However, CatBERTa predictions fail to capture this subtle energy difference, showing variations around 0.54 eV. Nevertheless, graph-assisted pretraining can help narrow the prediction range closer to the label energy range. After applying graph-assisted pretraining, the prediction ranges for that outlier cluster are reduced by 37\% and 48\% for the OC20 training and combined OC20 and OC20-Dense training cases respectively.

The outlier clusters on the right part of the plots are adsorption configurations of the \ce{CH} adsorbate and \ce{Ni16P16Se48} (0, 0, 1) catalyst pair. Systems containing \ce{CH} show an MAE of 0.40 eV for the GAP-CatBERTa\textsubscript{combined} case, which is not significantly greater than its overall MAE. Thus, this is not attributed to the adsorbates, unlike the previous case. For this cluster, four of the five points share the same configuration strings of [C P Se Se hollow [P Se Se C] [Se Ni Ni C] [Se Ni Ni C]]. It indicates a limitation of the current representation, which is less sophisticated than the graph representations. Further details about the duplicate configuration strings are provided in the supplementary information.

\newpage
\section{Prediction results of duplicate text sets}

\begin{figure*}[h!] 
\centering
\includegraphics[width=0.8\textwidth]{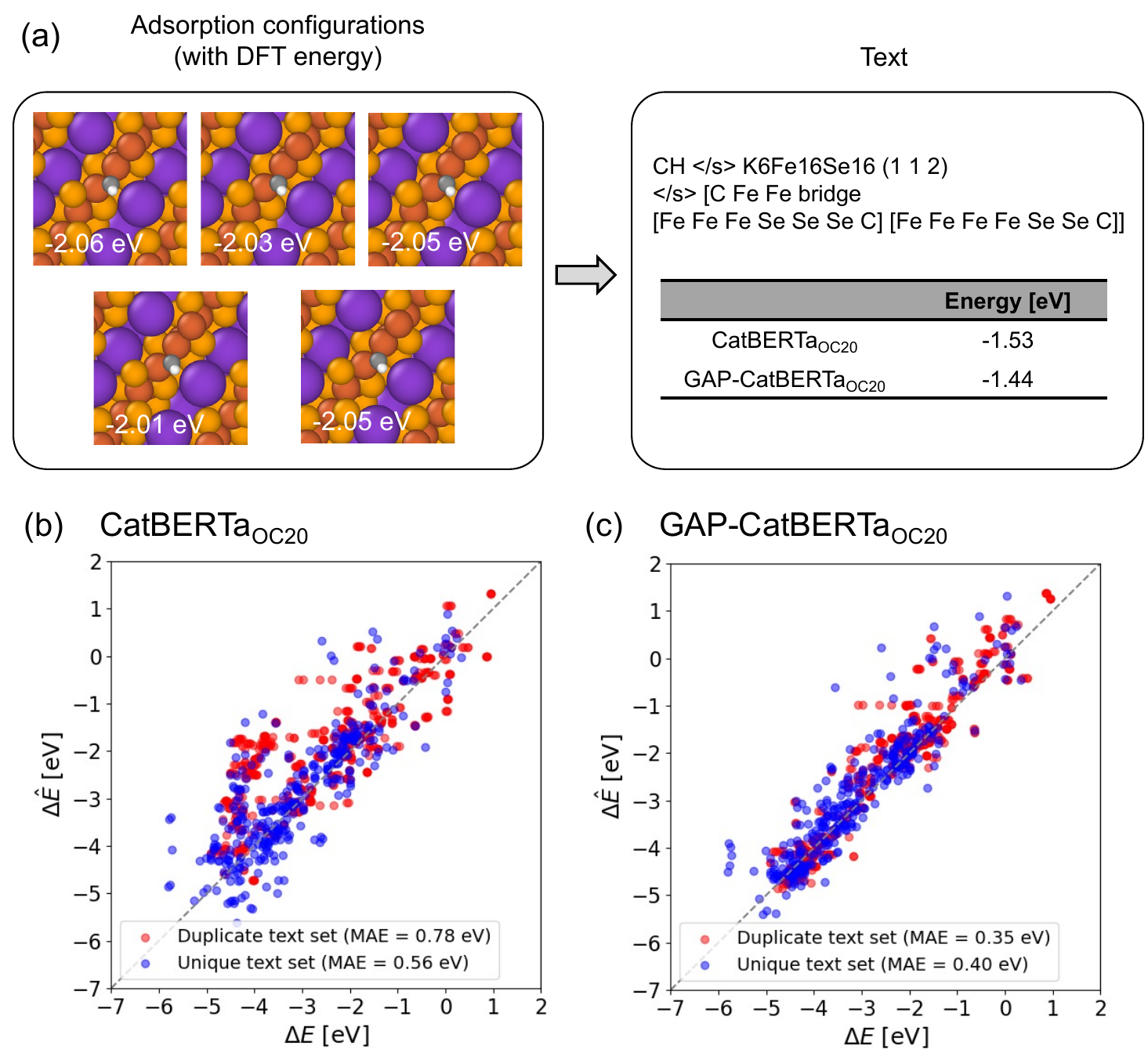} 
\caption{Comparative analysis of prediction performance on duplicate and unique text sets. \textbf{a} Instances illustrating the limitation of our textual approach in distinguishing minor structural differences. \textbf{b} MAE for each subset, comparing baseline scenarios to cases with enhancements applied.}
\label{fig:duplicates}
\end{figure*}

The textual representation used in this study, which does not fully encode the structure and atomic connectivity, is supposed to be less sophisticated than a graph representation in capturing structural nuances. This limitation is captured in the ``Accuracy improvement" section. This limitation is also illustrated in Figure \ref{fig:duplicates}(a), where the provided structures exhibit subtle differences that are difficult to discern visually. The graph representation successfully captures these nuances, leading to similar yet distinct configuration energy values for each structure, which range from -2.06 to -2.01 eV. In contrast, our textual representation does not discern these subtle differences, producing identical textual strings for all five structures. Consequently, this leads to identical energy predictions across these structures.

Duplicate texts representing different structures are one of the factors that impact the accuracy of predictions from the CatBERTa model. ``Duplicate text sets" refer to those that present the same text for different structures, while ``unique text sets" means those with no duplicates. As shown in Figure \ref{fig:duplicates}(b), in the CatBERTa\textsubscript{OC20} case, the duplicate text sets exhibit about 40\% higher MAE compared to the unique text sets. When graph-assisted pretraining is applied, the accuracy for the duplicate sets improves, along with the unique sets. This suggests that even though textual representations for duplicate sets do not capture subtle structural differences, their overall prediction accuracy can be improved by our pretraining strategy. This observation suggests that refining the textual string to generate distinct texts for similar adsorption configurations could improve CatBERTa's accuracy. 

Additionally, the issue of duplication can be resolved by incorporating additional feature sets that differentiate subtle configurations. In our previous paper, we demonstrated that multiple distinct feature sets could be included in the input data and processed within the same model framework. However, achieving high accuracy with this approach requires identifying feature sets that are highly correlated with the energy of the systems. This identification process is separate from developing a framework that connects graph embeddings and text embeddings. Once these powerful feature sets are identified, we can enhance the effectiveness of our text-based approach.

\end{document}